\documentclass[conf]{new-aiaa}
\pdfoutput=1  
\usepackage[utf8]{inputenc}

\usepackage[version=4]{mhchem}
\usepackage{siunitx}
\usepackage{longtable,tabularx}
\usepackage{graphicx,color}
\usepackage{epstopdf}
\usepackage{pgfplots}
\pgfplotsset{compat=newest}
\usetikzlibrary{plotmarks}
\usepackage{grffile}
\usepackage{amsmath}
\usepackage{natbib}

\usepackage{ulem}
\usepackage{color}

\usepackage{placeins}
\usepackage{textcomp}
\usepackage[euler]{textgreek}
\usepackage{xspace}
\usepackage{colortbl}
\usepackage[colorinlistoftodos]{todonotes}

\newcommand{\Detfull}{Det\textsubscript{full}\xspace} 
\newcommand{\Detpart}{Det\textsubscript{part}\xspace} 
\newcommand{\Def}{Def\xspace}
\newcommand{\drhodx}{$\frac{\partial \rho }{\partial x}$\xspace}
\newcommand{\drhody}{$\frac{\partial \rho }{\partial y}$ \xspace}

\newcommand{\tb}[2][black]{\textcolor{#1}{#2}}

\title{Investigation of the Exhaust Flow of a Pulse Detonation Combustor at different Operating Conditions based on High-Speed Schlieren and PIV}

\author{Mohammad Rezay Haghdoost\footnote{PhD Student, Laboratory for Flow Instabilities and Dynamics, Institute of Fluid Dynamics and Technical Acoustics, Technische Universität Berlin.}}
\affil{Technische Universität Berlin, D-10623, Germany}

\author{Daniel Edgington-Mitchell\footnote{Senior Lecturer, Laboratory of Turbulence in Aerospace and Combustion, Department of Mechanical and Aerospace Engineering, Monash University.}}
\affil{Monash University, Clayton, Victoria 3800, Australia}

\author{Christian Oliver Paschereit\footnote{Professor, Chair of Fluid Dynamics, Institute of Fluid Dynamics and Technical Acoustics, Technische Universität Berlin}}
\affil{Technische Universität Berlin, D-10623, Germany}

\author{Kilian Oberleithner\footnote{Professor, Laboratory for Flow Instabilities and Dynamics, Institute of Fluid Dynamics and Technical Acoustics, Technische Universität Berlin}}
\affil{Technische Universität Berlin, D-10623, Germany}

\begin{document}

\maketitle

\begin{abstract}
The exhaust flow of a Pulse Detonation Combustor (PDC) is investigated for different operating conditions. The PDC consists of two units, the deflagration to detonation transition section and the exhaust tube with a straight nozzle. High-speed high-resolution schlieren images visualize the shock dynamics downstream of the nozzle. The flow dynamics during one full PDC cycle is examined via high-speed Particle Image Velocimetry. A well-suited solid tracer particle for supersonic reactive flow is determined in a preliminary study to minimize the PIV measurement error.  The investigated operating conditions of the PDC differ in fill-fraction, which is the percentage of the tube filled with a reactive mixture. With increasing fill-fraction, the flow features grow in size and strength, as the propagation velocity of the leading shock increases. The blow down process of the PDC is characterized by several exhaust and suction phases. An increase in fill-fraction results in a stronger first exhaust phase, while the subsequent suction and exhaust phases remain almost unaffected.
\end{abstract}

\section{Nomenclature}

{\renewcommand\arraystretch{1.0}
\noindent\begin{longtable*}{@{}l @{\quad=\quad} l@{}}
D                                   & PDC exhaust tube diameter \\
ff                                  & fill-fraction \\
M                                   & flow Mach number \\
Ms                                  & shock Mach number \\
t                                   & time \\
u                                   & flow velocity \\
$u_{\mathrm{min}/\mathrm{max}}$     & velocity minima or maxima \\
$\dot{V}(t)$                        & volume flux \\
x,y,z                               & laboratory cartesian co-ordinate system \\
\textalpha                          & schlieren mirror offset angle \\
\textrho                            & density \\
\texttau\textsubscript{p}           & particle relaxation time \\
\textPhi                            & equivalence ratio \\
\end{longtable*}}

\section{Introduction} \label{intro}

Pressure Gain Combustion (PGC) is a well-established concept with the potential to drastically increase the efficiency of gas turbines.  Detonation based approaches have received significant attention in recent years, as even a small increase in total pressure across the combustor leads to a substantial increase in cycle efficiency. The thermodynamic benefit of the PGC technology is captured by the idealized Humphrey cycle, which replaces the isobaric Brayton cycle used to model conventional gas turbines. 

There are several approaches to realize PGC: Rotating Detonation Engine (RDE) ~\cite{kailasanath2017recent},  Resonant Pulse Combustor (RPC) ~\cite{yungster2016numerical} and Pulse Detonation Engine (PDE) ~\cite{pandey2016review,WOLANSKI2013125}, all of which have been the focus of research in recent decades.  Among the various  designs for PDE systems that have been proposed, the hybrid-PDE is one of the more promising configurations ~\cite{rasheed2011experimental}. Hybrid here refers to the integration of  an annular array of Pulse Detonation Combustor (PDC) tubes in a gas turbine engine, replacing the  conventional combustion chambers. 

The PDC cycle is composed of several phases. Initially, the detonation tube is filled with a detonatable mixture. The mixture is then ignited. A deflagration front propagates through the tube until it transitions to a detonation wave. Combustion products exit the tube after the detonation wave leaves the open end of the tube. Depending on the PDC design and operating condition none or several suction phases may occur. In the final phase, the tube is purged, enabling the reinitialization of the next cycle. 

One of the main challenges for implementing a PDC in a gas turbine is maintaining reliable operation of the turbine components. Excessively high temperatures, the presence of shocks, and high pressure and temperature fluctuations produced by the inherent unsteady combustion process are undesirable for both the compressor upstream and the turbine downstream of the PDC. In \tb{\citet{xisto2018efficiency} a  PDC-turbine system is investigated revealing that the mismatch between the transient inlet flow conditions of the rotor and the constant blade speed results in a significant amount of losses. In their study the PDC was directly attached to the turbine without any additional devices in between. If the PDC is not carefully integrated with the turbine components, these phenomena could easily eliminate any potential gain in cycle efficiency provided from PGC ~\cite{lisanti2017pulse}. One possible approach for coupling of PDCs and a downstream turbine would be using shape optimized  devices. Modifying the turbine blades to account for supersonic inlet flow is one possible approach \cite{paniagua2014design}. Also utilizing an additional device such as a plenum chamber between the PDCs and the turbine is currently the subject of research \cite{topalovic2019minimization}. The main purpose of this device is to minimize the turbine inlet flow fluctuations. However, to design such a device the detailed knowledge of the flow dynamics in the PDC exhaust is crucial.}

The basic feasibility of connecting a PDC to a conventional axial turbine has been demonstrated by different research groups. In the study performed by ~\citet{rasheed2011experimental} at the GE facility, eight PDC tubes were operated up to 30 Hz to drive a single stage turbine. Neither the coupling of the PDC with the turbine nor the turbine itself were optimized for the PDC application. Nevertheless, the PDC-fired operation showed a slight performance benefit when compared to the steady performance. In a study at the University of Cincinnati the performance of an axial turbine with six PDC tubes was investigated. ~\citet{glaser2007performance} conducted performance measurements for both a constant pressure combustor driven turbine and for the PDC-turbine case. It was found that the performance of the PDC driven turbine was comparable to that of a constant pressure combustor driven turbine across its operating map. They further show that an increase in fill-fraction, which is the percentage of the tube filled with a reactive mixture, causes a decrease in turbine efficiency. The mechanism responsible for this decrease in efficiency was not determined.  
 
To develop an efficient hybrid PDE, the coupling of the PDC and the turbine needs to be understood in detail. For this purpose, the knowledge of the highly unsteady exhaust flow of the PDC tube  is crucial. This has been the subject of a number of research efforts in the last decades \cite{arienti2005numerical,schultz2000detonation,allgood2003computational,ma2005thrust,hoke2009schlieren,opalski2005detonation}. Among these, \citet{allgood2003computational} investigated the blow down process of an overfilled PDC tube using a shadowgraph technique. They visualized the dynamic evolution of the exhaust flow for straight and convergent nozzles. \citet{glaser2004experimental} performed similar shadowgraph visualizations for different equivalence ratios and fill fractions. Despite  limited image resolution, they were able to detect some of the flow features, such as small vortices and the corresponding slipstream within the primary vortex ring. \citet{opalski2005detonation} studied the exhaust flow of an overfilled PDC quantitatively via Particle Image Velocimetry (PIV). They characterized the transient flow field based on ensemble-averaged velocity fields. Their data shows a short duration of a higher velocity outlet flow of about 4 ms, where a peak axial velocity along the tube axis centerline of 1880 m/s was measured.  Overall, these previously conducted expeirments emphasize the highly unsteady nature of the flow field exiting the PDC  tubes which poses significant challenges on experimental techniques to reveal the flow physics involved.

With the aim to manipulate the unsteadiness of the PDC exhaust flow and to improve the overall operability, different parameters and geometrical variations of the PDC have been investigated.  By utilizing an ejector downstream of the PDC \citet{opalski2005detonation} were able to reduce the unsteadiness of the exhaust flow successfully for an overfilled configuration. \citet{allgood2005experimental}  observed the performance of the ejector used in their study to be sensitive to the inlet geometry as well as its axial position relative to the exhaust plane of the PDC.  
Moreover, the impact of the fill-fraction on the PDE performance has been the subject of research in some detail  \cite{allgood2003computational,li2000numerical,schauer2001detonation}. Thereby, an increase in fuel specific impulse with smaller fill-fraction was shown numerically \cite{li2000numerical} and experimentally \cite{schauer2001detonation}. Although there are some performance measurements for different fill-fractions and a limited number of studies on the exhaust flow of the PDC, a detailed experimental study on the impact of fill-fraction on the exhaust flow is missing. Moreover,  very little attention has been paid to the impact of the fill-fraction on the evolution of the exhaust flow, which is important for the coupling with the turbine. 
 
\tb{In the current study a series of schlieren and PIV measurements are conducted at the outlet of a PDC for various operating conditions. The  operating conditions are determined primarily by the fill-fraction, which is one of the main controlling parameters for the PDC cycle. The recorded high-resolution high-speed data allow for a detailed investigation of the exhaust flow. The shock dynamics at the initial stage of the PDC cycle are characterized in detail based on schlieren images, while the the exhaust flow rate is quantified via PIV for the full PDC cycle. The combination of these two measurement techniques allows the important local and global features of the exhaust flow to be tracked for different fill fractions, which provides a sound empirical base for future numerical and analytic studies. E.g., this study provides experimental proof for substantial flow reversal during the PDC cycle which is of significant importance for coupling of the PDC with a downstream turbine.}

\section{Methodology} \label{setup}

\FloatBarrier
\subsection{PDC test-rig and experimental setup}\label{sec:exp_setup}
\begin{figure}
	\centering
	\includegraphics[width=1\textwidth]{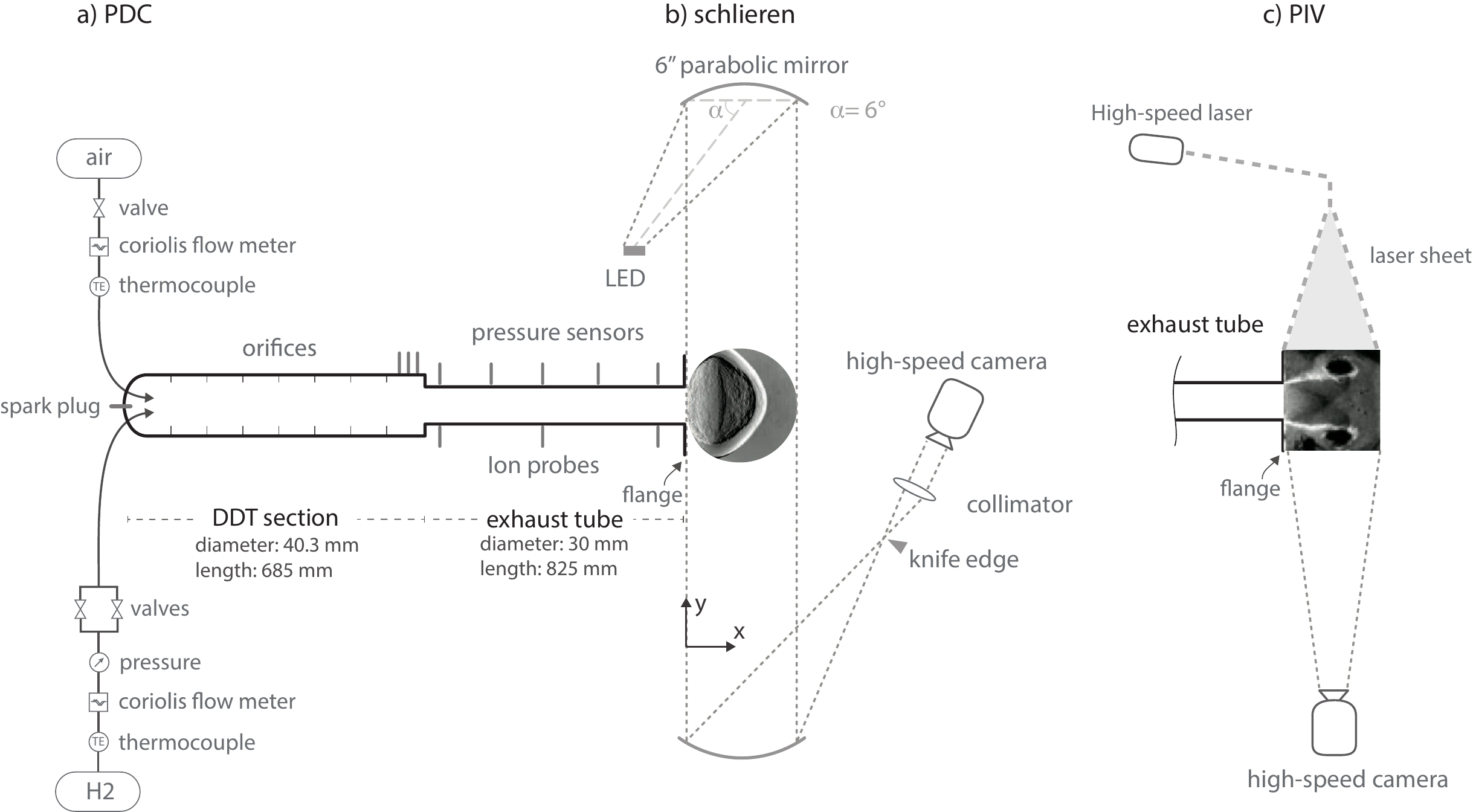}
	\caption{Illustration of the experimental setup showing a) the Pulse Detonation Combustor, b) the high-speed schlieren setup, c) the high-speed PIV setup at the tube exit. The images at the tube exit indicate the enlarged measurement domain.}
	\label{fig:experimental_setup}
\end{figure}

Figure~\ref{fig:experimental_setup} presents a schematic of the Pulse Detonation Combustor (PDC) and the instrumentation of the experimental setup. The PDC used in this study consist of two sections: the section where the Deflagration to Detonation Transition (DDT) takes place and the exhaust tube. In this valveless design, the air supply is not being modulated but is attached directly to the upstream end of the PDC tube .  Hydrogen is injected through eight circumferentially-distributed fuel lines. The design of the air and hydrogen injection scheme is described in~\citet{gray2017compact}.

Combustion is initiated with a spark plug, positioned at the upstream end of the DDT section. Orifices positioned in the DDT section accelerate the flame propagation. The speed of the incident shock is determined by five piezoelectric pressure probes (PCB112A05), flush-mounted to the exhaust tube, as shown in figure~\ref{fig:experimental_setup}(a). To assess whether the detonation state has been reached before the wavefront enters the exhaust tube, three additional closely-spaced pressure probes are placed at the rear of the DDT section.

The combustion front is tracked by three flush-mounted ionization probes, mounted on the opposite side of the pressure probes. The  ionization probes, fabricated in-house, consist of two electrodes separated by a ceramic coating. The ionized species in the combustion region allow an electric current to flow, as a potential difference is applied to the electrodes. The resulting voltage drop indicates the arrival of the reaction front. 

The data from pressure and ionization probes is acquired simultaneously on 11 channels, using a National Instruments MXI-Express DAQ system at 1 MHz sampling rate. The mass flow rate of air and hydrogen is measured with two different Endress + Hauser Coriolis mass flow meters. Two type K thermocouples are used for measuring the temperature of air and hydrogen. A Festo pressure transmitter (SPTW-P10R) is used to measure the pressure of hydrogen upstream of the injection valves.

As shown in figure~\ref{fig:experimental_setup}(b), a standard z-type schlieren setup is used with two 6-inch parabolic f/8 mirrors for collimating and refocusing of light. The schlieren images are captured with a 1-megapixel Photron SA-Z high-speed camera at frequencies up to \mbox{80 kHz}. A pulsed LED is used as a light source, as suggested by~\citet{willert2012assessment}. The overdriven-operated \tb{Luminus LED PT-120-TE} at 10.3 voltage  provides a high intensity light pulse with an exposure time of 1 \textmu s.

Particle images are acquired using the SA-Z camera \tb{with an array size of 1024 × 1024 pixels } mounted orthogonally to the PDC outlet as illustrated in figure~\ref{fig:experimental_setup}(c). A Darwin-Duo diode-pumped Nd:YLF laser operated at its maximum frequency of 10 kHz supplies illumination. A series of lenses create a light sheet of approximately 1 mm width at the center of the exhaust tube. An ILA 5150 synchronizer is used for timing of the laser and the  camera. An air-driven fluidized bed PIVsolid 8 is used to seed both the tube and the ambient air close to the PDC outlet. Uniform distribution of the ambient seeding is aided by using two FDX fluidic oscillators \cite{bobusch2013numerical}. The exhaust tube used for the PIV measurements is 45 mm longer than the one shown in figure~\ref{fig:experimental_setup}(a), which is used for the schlieren measurements.

\tb{A laser pulse separation of 4 $\mu$s is chosen as a compromise for small and large flow velocities during the entire cycle. The magnification of 13.75 px/mm results in a particle displacement of 5.5 px for 100 m/s and 105 px for 1900 m/s. In order to account for the relatively large particle displacement at high velocities an iterative multi-grid approach including image deformation with an initial sampling window of 128x128 px is chosen \cite{willert1991digital,scarano2001iterative}. Erroneous velocity vectors are identified by a dynamic mean value operator and replaced by interpolation with immediate neighbours. For the applied evaluation strategy the sub-pixels error is in the range of 0.1 pixel based on the particle image diameter \cite{raffel2018particle}. This corresponds to a measurement uncertainty of 1.8 m/s. The error is approximately between 0.01\% and 6\% of the maximum velocity during the entire exhaust phase.}

\definecolor{dunkelgrau}{rgb}{0.87,0.87,0.87}
\definecolor{hellgrau}{rgb}{0.93,0.93,0.93}
\definecolor{sehrhellgrau}{rgb}{0.97,0.97,0.97}

\begin{table}
		\centering
	\begin{small}
		\caption{PIV parameters. \label{tab:PIV_para}}
		\begin{tabular}{lc} 
			\rowcolor{sehrhellgrau}
			IW [px]                                   &            $32 \times 24$           \\    
			\rowcolor{hellgrau}
			IW [D]                                   &            $0.08 \times 0.06  $       \\  

			\rowcolor{sehrhellgrau}
			Overlap     [\%]                       &             50                 \\     
			\rowcolor{hellgrau}
		 
			\tb{Pulse distance       [\textmu s ]}                  &                   \tb{4}     \\ 
			
			\rowcolor{sehrhellgrau}
			\tb{Pulse width [ns]} & \tb{550 - 600} \\
			
			\rowcolor{hellgrau}
			\tb{Pulse power [W]} & \tb{40 - 45} \\
			
			\rowcolor{sehrhellgrau}
			Field of view  [D]						      &                    $2.3 \times 2.5$      \\     	
			\rowcolor{hellgrau}
			Digital resolution   [px/D]                  &                   411       \\         
			\rowcolor{sehrhellgrau}
			Particle relaxation time [\textmu  s]          &                    0.84        \\                 
			\rowcolor{hellgrau}
			\tb{Exposure time   [ns]}                      &                   \tb{159}       \\
		\end{tabular}
	\end{small}
\end{table}

\FloatBarrier
\subsection{Preliminary investigation of PIV seeding materials}\label{sec:seeding}

The flow-tracking fidelity of the tracer particle is a fundamental assumption for tracer particle-based measurement techniques such as PIV. This becomes critical for detonation applications, due to the characteristics of reactive supersonic flows. On the one hand, the fluid velocity changes abruptly across strong shock waves. On the other hand, the high temperatures associated with reactive flows precludes the use of liquid seeding material, necessitating the use of oxidized metals with higher mass density than the surrounding fluid. Hence, high-inertia solid particles crossing shock waves may result in a significant slip velocity and therefore biased measurement results. These errors may be further exacerbated by a non-uniform  particle size distribution  \cite{mitchell2011particle}. The fundamental particle size for solid particle seeding can be misleading; particles at the micro- and nano-scale will form larger agglomerates, both at rest and in flight \cite{yao2002fluidization}. Strong shear forces in a compressible flow can then break these agglomerates up in-flight, resulting in a large range of particle scales, both too small to be well-resolved by the optical system, and too large to accurately track the flow velocity. Hence, a well-suited seeding material is required to minimize the measurement error. To this end, the particle response to a step change in fluid velocity and the raw image quality of six different seeding particles have been investigated in a preliminary study, which will be outlined in the following. Additional information can be found in ref.~\cite{HaghdoostTracer}.

To investigate the response time of different particles, PIV measurements with different seeding materials such as TiO\textsubscript{2}, SiO\textsubscript{2} and ZrO\textsubscript{2} were conducted for the velocity step across a Mach disk of a highly-underexpanded steady jet in a separate facility. In addition, schlieren measurements were conducted to obtain the actual location of the Mach disk. By comparing the PIV results with the schlieren images, the particle lag in the presence of shock waves is evaluated.

\begin{figure}
	\centering
	\includegraphics[width=.8\textwidth]{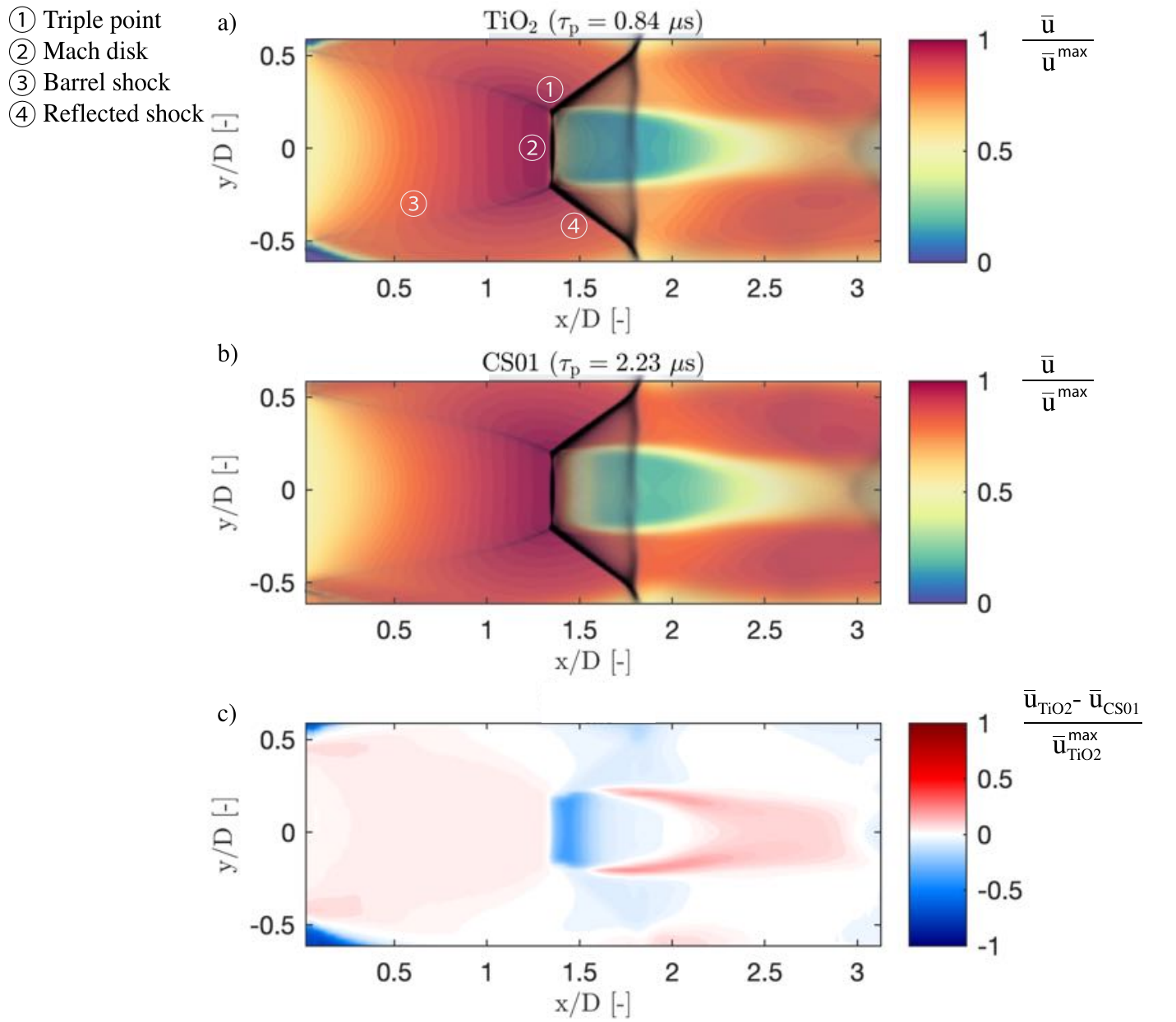}
	\caption{Steady underexpanded jet with a nozzle pressure ratio of 5.2. Gray scale contours: schlieren image intensity; colour contours: axial velocity determined from PIV-Data. (a) PIV with TiO\textsubscript{2} seeding (b) PIV with ZrO2CS01 seeding and (c) their normalized deviation. The spatial coordinates are normalized with respect to the nozzle diameter D. }
	\label{fig:overlay_comp}
\end{figure}

Figure~\ref{fig:overlay_comp} presents the schlieren and PIV results for two different seeding materials. The intensity of the schlieren image is inverted and overlaid on the velocity contour plot gained from the PIV measurements. The velocity $\bar{u}$ corresponds to an average of 5000 snapshots and is normalized by the peak velocity $\bar{u}_{max}$.  The comparison of the schlieren and velocity field for the TiO\textsubscript{2} particle (figure~\ref{fig:overlay_comp}(a)) shows very good agreement between the locations of the Mach disk, reflected shock and barrel shock. The same plot for the PIV data using the zirconium dioxide ZRO2CS01 as seeding is presented in figure~\ref{fig:overlay_comp}(b).  The slow particle response leads to smearing of the velocity gradients, particularly notable in the vicinity of the Mach disk.  \tb{A quantitative comparison between the two velocity fields is given in figure~\ref{fig:overlay_comp}(c). The largest deviations occur in areas where the velocity gradients are the largest. The TiO\textsubscript{2} particles move at a higher speed in regions of positive velocity gradient. This corresponds to the red region in \mbox{0 < x/D < 1.34} in figure~\ref{fig:overlay_comp}(c), where an expansion fan forms at the nozzle lip. Downstream of the Mach disk the velocity decreases abruptly. Hence, the largest discrepancy between the particle velocities appears in this region at x/D $\approx$ 1.43. Further downstream there is an additional red area   \mbox{x/D > 1.55}  \& \mbox{-0.25 < y/D < 0.25}. This indicates that both, the vortices in the slipstream and the acceleration of the flow due to the reflection of the reflected shock at the jet boundary as expansion fan are captured more accurately with the TiO\textsubscript{2} particles. These results clearly emphasize that by using seeding with slow particle response, the strong velocity gradients in supersonic flows cannot be accurately measured with the PIV technique.}

\begin{figure}
	\centering
	\includegraphics[width=1\textwidth]{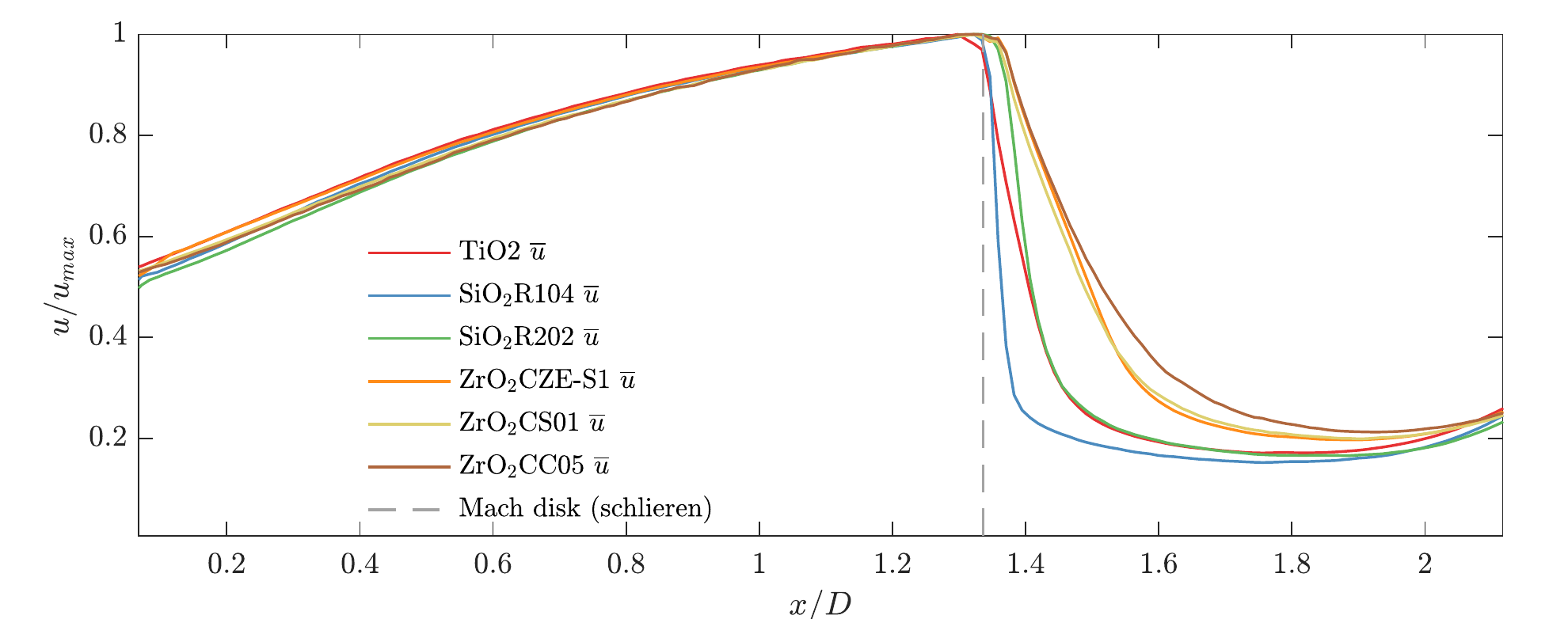}
	\caption{Axial profiles of normalized mean velocity along the jet centerline for six different tracer materials.}
	\label{fig:velProfile_NPR52}
\end{figure}

 Figure~\ref{fig:velProfile_NPR52} presents the axial mean velocity along the jet centerline, normalized by the maximum velocity. A strong decay of the velocity after crossing the Mach disk is noticeable for all investigated seeding materials.  The slope of the  curves, however, indicate the different response of the seeding materials to the step change in velocity downstream of the Mach disk. The SiO\textsubscript{2}R104 exhibits the best response among the investigated seeding materials followed by the SiO\textsubscript{2}R202 and TiO\textsubscript{2}. However, all three zirconium dioxide materials show a longer region of deceleration after passing the shock wave.

The particle response is typically quantified by the particle relaxation time \texttau\textsubscript{p}. This is the time required for the velocity-lag, downstream of a step change in velocity, to be reduced by the factor 1/e. The particle response time  for each seeding material is evaluated by examining their motion through the Mach disk using the method developed by \citet{melling1997tracer} and \citet{ragni2011particle}. The determined values range from 0.32 to 2.29 \textmu s as shown in table~\ref{tab:RelaxationTime}. Best particle response is achieved using SiO\textsubscript{2} particles followed by the TiO\textsubscript{2} particles.

\begin{table}
		\centering
	\begin{small}
		\caption{Measured relaxation time for tracer materials. \label{tab:RelaxationTime}}
		\begin{tabular}{lc} 
		    \rowcolor{dunkelgrau}
		    Name                        & \texttau\textsubscript{p} in \textmu s     \\
			\rowcolor{sehrhellgrau}
			SiO\textsubscript{2}R104    &    0.32   \\    
			\rowcolor{hellgrau}
			SiO\textsubscript{2}R202    &    0.61   \\ 
			\rowcolor{sehrhellgrau}
			TiO\textsubscript{2}        &    0.84   \\ 
			\rowcolor{hellgrau}
			ZrO\textsubscript{2}CC05    &    2.22   \\ 
			\rowcolor{sehrhellgrau}
			ZrO\textsubscript{2}CS01    &    2.23   \\ 	
			\rowcolor{hellgrau}
			ZrO\textsubscript{2}CZE-S1  &    2.29   \\ 	
		\end{tabular}
	\end{small}
\end{table}

\begin{figure}
	\centering
	\includegraphics[width=1\textwidth]{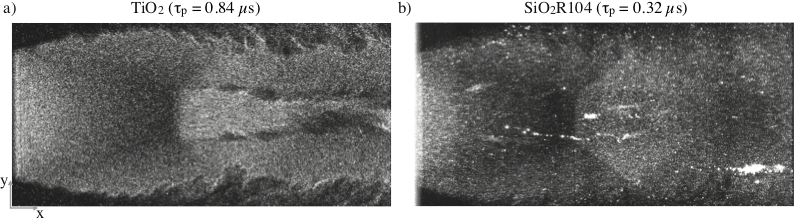}
	\caption{Mie scattering images of SiO\textsubscript{2}R104 and TiO\textsubscript{2} for a steady underexpanded jet.}
	\label{fig:raw_images}
\end{figure}

The raw image quality of the PIV images has also been taken into account for choosing well-suited seeding particles.  Inspection of the raw images  show that the TiO\textsubscript{2} powders exhibit more uniform seeding than the other materials. Figure~\ref{fig:raw_images} shows exemplary snapshots of the raw images for TiO\textsubscript{2} and SiO\textsubscript{2}R104. The silicon dioxides exhibit a much wider range of particle intensities in the images. Moreover, the TiO\textsubscript{2} raw images exhibit better contrast. Considering all these aspects, the TiO\textsubscript{2} shows an overall best performance. Therefore, the TiO\textsubscript{2} particles has been chosen as seeding particle for the PIV measurement of the PDC.

\FloatBarrier
\subsection{PDC operating conditions \label{sec:Operating conditions}}
The present experimental investigations are conducted in a single-cycle manner, i.e. only a single combustion event (deflagration or detonation) is undertaken for each measurement. The tube is first filled with a combustible mixture for a certain amount of time and subsequently ignited. The corresponding exhaust flow is captured with schlieren and PIV techniques. All cases considered in this study differ only in the hydrogen filling time. This leads to different  fill-fraction and equivalence ratio, as described below. 

\begin{figure}
	\centering
	\includegraphics[width=1\textwidth]{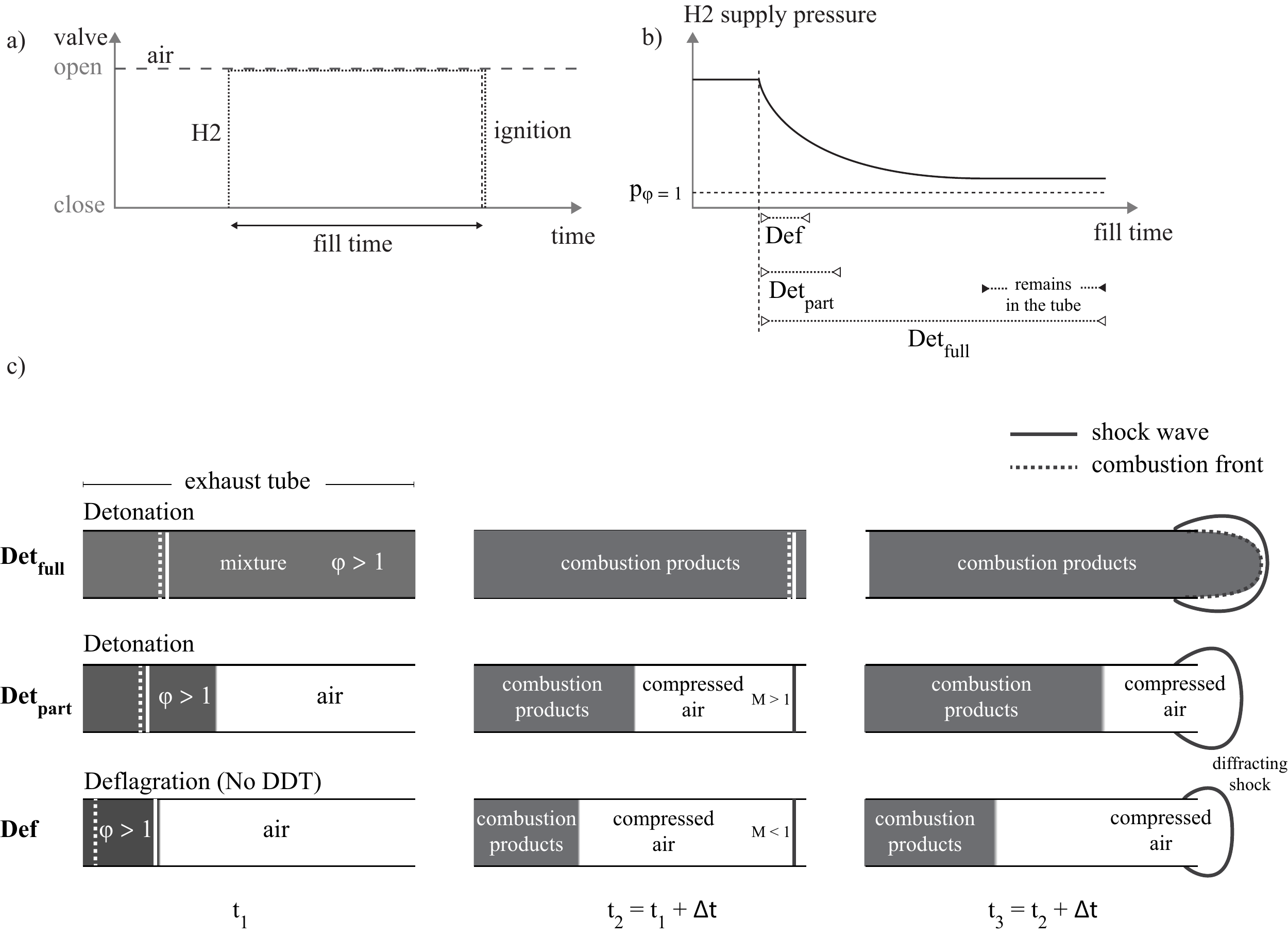}
	\caption{a) Valve switching timeline for injection of air and hydrogen into the test rig. b) Supply pressure of \mbox{H\textsubscript{2}} as a function of filling time is shown schematically for different cases. c) Illustration of the experimental cases considered in this study.  Different fill-fraction (ff) and equivalence ratio (\textPhi) in the exhaust tube are shown as hydrogen filling time changes:
		\mbox{\textPhi \textsubscript{Def} \textgreater\
		\textPhi \textsubscript{\Detpart} \textgreater\
		\textPhi \textsubscript{\Detfull};}
		ff \textsubscript{Def} \textless\
		ff \textsubscript{\Detpart} \textless\ 
		1  \textless\ 
		ff \textsubscript{\Detfull}. The propagation of the shock and reaction front within and outside of the exhaust tube is represented schematically for each case.
	}
	\label{fig:cases_and_valve_time_line}
\end{figure}

A schematic illustration of the valve timing during the fill process is shown in figure~\ref{fig:cases_and_valve_time_line}(a). In this valveless design the pressure in the air supply line is chosen so that during the filling time the air flows continuously at 100 $\frac{kg}{h}$. As indicated in figure~\ref{fig:cases_and_valve_time_line}(a), the filling time refers to the time period during which hydrogen is injected into the PDC tube rig. The ignition is initiated at the same time as the hydrogen valve is closed. 

Changing the filling time results in different operating conditions for the PDC, with low filling time corresponding to  low fill-fraction. In our setup, it leads also to a gradient in equivalence ratio \textPhi\ along the tube. The dependence of equivalence ratio on the filling time is due to the fact that the pressure in the \mbox{H\textsubscript{2}} supply line decreases exponentially once the valve is opened. This is illustrated in figure~\ref{fig:cases_and_valve_time_line}(b). The pressure reaches a plateau value, approximately one second after the valve is opened. This pressure level was set to achieve the desired mass flow rate corresponding to \mbox{\textPhi\ $>$ 1}. Hence a decreasing supply  pressure  during the filling process results in a stratification of the mixture with a positive equivalence ratio gradient along the tube. 

Three different cases with various filling times are presented in this paper, referred to as \Detfull, \Detpart and Def. A schematic representation of these cases is given in figure~\ref{fig:cases_and_valve_time_line}(c) at three different time stages of the PDC cycle. A rich mixture is chosen for all operating conditions and the fill-fraction is varied to investigate its impact on the exhaust flow.
In case of \Detfull, the filling time is roughly 200 times longer than it needs to fill the whole tube. Thus not only the exhaust tube is fully filled with the detonable mixture but the mixture also extends downstream of the tube exit (ff $>$ 1). In this case, the hydrogen mass flow needed for a rich mixture is adjusted by setting the pressure in the hydrogen supply line. A comparison of the CJ detonation velocity computed with the NASA CEA code~\citep{CEA_Nasa} and the one determined with the ionization and pressure probes shows that the tube is filled with a rich mixture corresponding to \textPhi $\approx 1.4-2$. 

\begin{figure}
	\centering
	\includegraphics[width=0.8\textwidth]{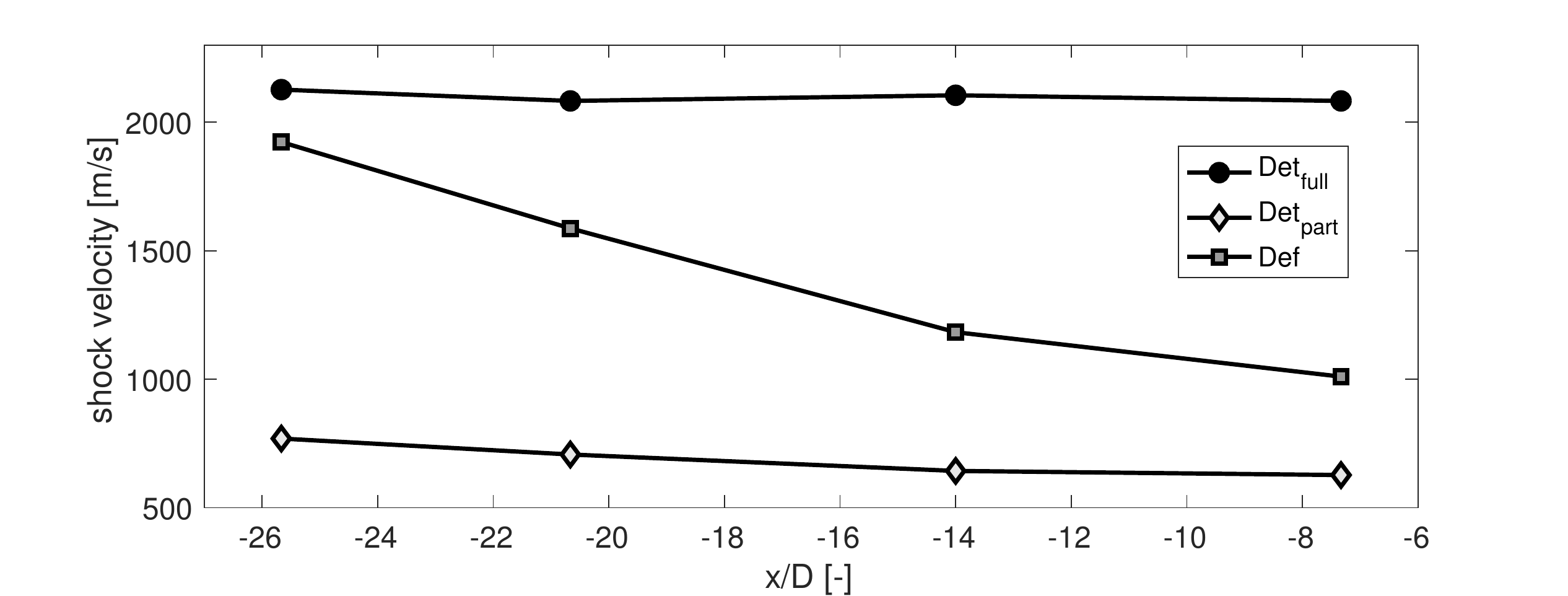}
	\caption{\tb{Leading shock velocity in the laboratory reference frame based on time-of-flight method.}}
	\label{fig:shock_velocity}
\end{figure}

\tb{As illustrated in figure~\ref{fig:cases_and_valve_time_line}(c), the detonation wave entering the exhaust tube leaves the tube also as a detonation wave.  However, this is not the case for the two other investigated operating conditions.  A shorter filling time is chosen for the \Detpart case when compared to the \Detfull case.} Therefore, the exhaust tube of the PDC is filled only partially with a reactive mixture (ff \textless 1). As illustrated in figure~\ref{fig:cases_and_valve_time_line}(c) at the time t\textsubscript{2}, the detonation wave is transmitted as a shock at the contact surface (mixture-air-interface) and moves downstream through the air; the data from the pressure and the ionization probes confirm that the shock wave leaves the tube without a reaction zone directly behind it, as illustrated in t3 in figure~\ref{fig:cases_and_valve_time_line}. \tb{Following the collision of the detonation wave with the contact surface, also a rarefaction wave is generated at the contact surface, which propagates toward the upstream end of the PDC \cite{li2003partial}. While the transmitted shock wave compress and accelerate the air toward the tube exit, the detonation products are expanded and accelerated toward the tube exit by the the rarefaction wave.}

\tb{A CJ detonation wave is followed by a series of rarefaction waves (Taylor waves), which decelerate the burned gas to satisfy the closed wall boundary condition at the upstream end of the PDC \cite{taylor1950dynamics}. Once the detonation wave is transmitted as a shock wave at the contact surface, the Taylor waves overtake the transmitted shock wave. This results in an considerable attenuation of the shock wave as it propagates through the air \cite{peace2018detonation}. Similar to a shock tube also viscous effects contributes to the deceleration of the leading shock wave \cite{mirels1957attenuation,ii1955theoretical}.} 

\tb{These considerations are in line with shock propagation velocity as shown in Figure \ref{fig:shock_velocity}.  The velocities were estimated from the experiments using the time-of-flight method. The arrival time of the shock wave at the location of the pressure probes was determined from the pressure signals, and the shock velocity between the probes was estimated by using the arrival time and the distance between the pressure probes. The pressure probes in the exhaust tube are located at x/D = -27, -24, -17, -11 and -4. According to figure \ref{fig:shock_velocity}, the detonation wave propagates at a nearly constant velocity  along the entire exhaust tube in the \Detfull case, while a significant deceleration of the shock propagation velocity is evident for the \Detpart case.}

For the \Def case an even shorter filling time is chosen (see illustrations in fig~\ref{fig:cases_and_valve_time_line}) and no transition to detonation takes place. In comparison to the \Detpart case the  mixture is richer and fills even less volume of the exhaust tube. \tb{An accelerated deflagration wave propagates toward the tube exit until the combustion front quenches at the mixture-air-interface. Due to the closed wall boundary condition at the upstream end of the PDC, the increase in the specific volume of the combustion products results in a displacement of the reactants ahead of the combustion front. Therefore, the reactants move toward the open-end of the tube prior to the combustion. As a result of the displacement of the flow, precursor compression waves are formed in front of the reaction front \cite{lee2008detonation}. Rapid increase of the flame surface is caused by the obstacles in the DDT section. Furthermore, the flame burning rate increases due to a number of instabilities such as Kelvin-Helmholtz and Rayleigh-Taylor instability \cite{ciccarelli2008flame}. Further pressure waves are generated ahead of the flame front, as the flame propagates along the DDT section, while its burning rate increases. These pressure waves eventually coalesce to a single shock wave leaving the PDC. However, due to the low fill fraction the accelerated flame front does not catch up with the preceding shock wave to form a detonation wave.  Also for the \Def case, the measured pressure next to the tube exit indicates an expansion wave attached to the leading shock wave. This may be caused by a highly accelerated flame, which results in expansion wave to satisfy the closed wall boundary condition \cite{ciccarelli2008flame}. The expansion waves and viscous effects decelerate the leading shock wave, while it propagates along the exhaust tube (figure \ref{fig:shock_velocity})}. 

The Mach number of the incident shock is the key quantity for the flow field behind it. In particular the initial flow evolution at the tube exit is governed by the shock strength at the tube exit.  Therefore, the incident shock velocity at the tube exit is determined by extrapolating the estimated shock velocity along the tube. The shock velocities in the laboratory reference frame at the tube outlet for all the operating conditions are listed in table~\ref{tab:velocity}. \tb{Also the shock Mach number is given in  table~\ref{tab:velocity}. According to Rankine-Hugoniot equations the flow behind a non-reactive 1D shock wave is supersonic if the shock strength is beyond Ms = 2.07. Thus, at the tube exit (x/D = 0) the flow behind the incident shock wave is subsonic for the \Def and supersonic for the \Detpart case. For the sake of completeness, the shock Mach number for the \Detfull case in reference of the unburned gas is also given in table~\ref{tab:velocity}. However, the flow behind the reactive CJ detonation is sonic relative to the wave but subsonic in laboratory reference frame.} PIV measurements were performed only for the \Detfull and \Def cases. As shown in table~\ref{tab:velocity} the shock wave velocity is somewhat reduced when PIV is conducted. This is due to the additional seeding air required for PIV, which dilutes the mixture. 

\definecolor{dunkelgrau}{rgb}{0.87,0.87,0.87}
\definecolor{hellgrau}{rgb}{0.93,0.93,0.93}
\definecolor{sehrhellgrau}{rgb}{0.97,0.97,0.97}

\begin{table}
	\centering
	\caption{Velocity and Machnumber of the incident shock at the tube exit.\label{tab:velocity}}
	\begin{tabular}{ccccc} 
		\rowcolor{dunkelgrau}
		operating condition	& u\textsubscript{schlieren} [m/s] & u\textsubscript{PIV}[m/s] 	&	\tb{ Ms\textsubscript{schlieren} [-]}	& \tb{ Ms\textsubscript{PIV}[-] }\\
		\rowcolor{sehrhellgrau}
		\Detfull		  &			2058 &  1930 &\tb{ 4.88} &\tb{  4.78} \\ 		
		\rowcolor{hellgrau}
		\Detpart		&		909 &  - & \tb{ 2.53} &	- \\        
		\rowcolor{sehrhellgrau}
		\Def				&	606	& 508 & \tb{ 1.65} &\tb{  1.37} \\ 			
\end{tabular}
\end{table}

\FloatBarrier

\section{Influence of fill-fraction on initial jet evolution} \label{sec:jet_evolution_initial_stage}
The dynamic evolution  of the PDC exhaust flow depends on its operating conditions. In case of \Detfull, the incident shock exits the tube as a detonation wave, while for the \Detpart cases the detonation wave dies out before it exits the tube and the incident shock is separated from the reaction zone. This is analog to the \Def case where a single shock exits the tube. In the following, the evolution of the incident shock and other shock-dynamic characteristics of the exhaust flow at its initial stage are investigated in detail, based on the acquired high-speed schlieren images. 

\subsection{\textbf{Small fill-fraction (\Def)}}
We start with the discussion of the \Def case, as the flow visualizations are the easiest to interpret, and many of the features are common to all three cases.  While the fill-fraction is smallest for this case, the time span between the shock arrival and the contamination of the schlieren images by the combustion products is largest.   

\begin{figure}
	\centering
	\includegraphics[width=0.7\textwidth]{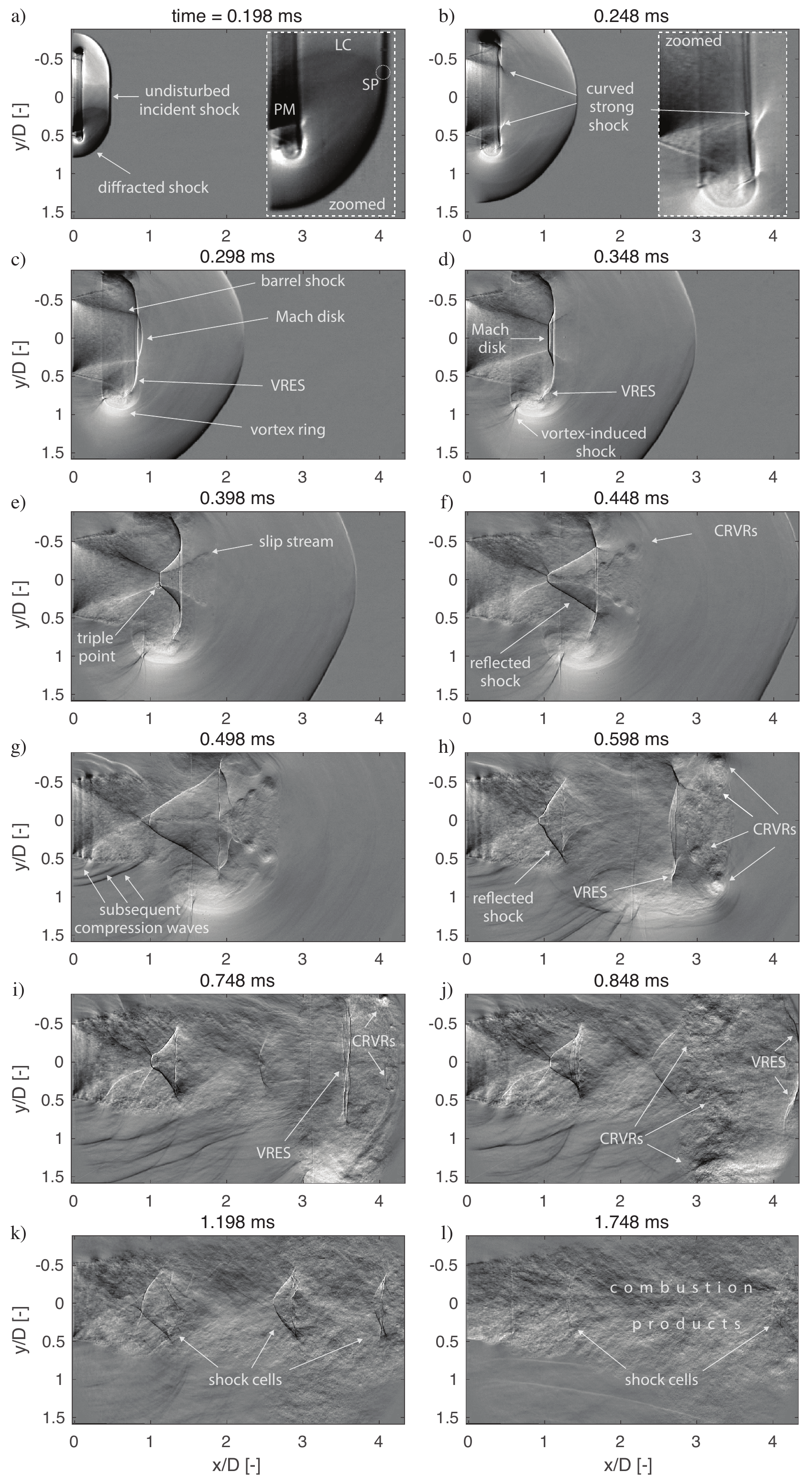}
	\caption{Detailed evolution of the exhaust flow at its early stage. The appearance and further development of dominant flow features is visible. The given time above each image corresponds to the time elapsed after the shock passed the pressure probe next to the exhaust tube outlet. Case: Def.}
	\label{fig:Jet_evolution_Def}
\end{figure}

The corresponding schlieren images of the \Def case are presented in figure~\ref{fig:Jet_evolution_Def}. These 12 \drhody images depict the key moments of the flow evolution from the time the shock exits the PDC tube until the the density gradients of the succeeding combustion products dominate the schlieren images. In all images, the primary exhaust flow moves from left to right. The x- and y-axis are normalized by the tube diameter D = 30 mm, and the origin of the axis corresponds to the point on the tube centerline at the tube exit. The  time shown above the image corresponds to the time after the shock passes the pressure sensor, which is mounted 4D upstream of the tube exit. The first seven images are uniformly separated by a time interval of 50 \textmu s, while the remaining images are separated by longer intervals. 

From the time series of the schlieren images presented in  figure~\ref{fig:Jet_evolution_Def}, the flow evolution can be determined in detail.  Figure~\ref{fig:Jet_evolution_Def}(a) shows the moment right after the shock diffracts around the tube exit corner.  At this time instant a part of the shock has already undergone an \tb{axisymmetric} diffraction as indicated by the partially curved shock. Both the undisturbed and the diffracted shock are clearly visible. \tb{An unsteady expansion wave (EW) moves toward the tube and accelerates the subsonic flow inside the tube. The exhaust flow expands further and forms a Prandtl-Meyer expansion fan (PM) centered at the tube exit corner \cite{sun1997formation}.} The unsteady expansion wave (EW) marks the separation point (SP) between the undisturbed incident shock and the diffracted shock.

Figure~\ref{fig:Jet_evolution_Def}(b) depicts the moment when a slightly curved strong shock is being formed at the outer region of the jet next to the barrel shock. Downstream of the expansion fan, the pressure is lower and the velocity is higher compared to the flow being set into motion by the incident shock. Hence, a system of shocks consisting of a Mach disk and a barrel shock occurs. This shock system process the exhaust flow to match the pressure and velocity of the over-expanded exhaust flow with the one being set into motion by the incident shock \cite{friedman1961simplified}. The upper and  lower curved strong shocks, marked in figure~\ref{fig:Jet_evolution_Def}(b) propagate toward the jet centerline to form the Mach disk. When these shocks coalesce, a Mach disk is formed, which develops to its typical disk shape shortly after, as can be seen in figure~\ref{fig:Jet_evolution_Def}(c). 

A characteristic feature of the PDC exhaust flow is the vortex ring which arises as the shear layer at the trailing edge rolls up. There are two vortex ring-associated shocks, the vortex-induced shock and the  vortex-ring-embedded shock (VRES), which are marked in figures~\ref{fig:Jet_evolution_Def}(c)-(d). The VRES exist only within the vortex ring and not inside the jet core. However, a part of this shock appears in the schlieren image as a vertical line along the entire jet core. This is due to the fact that the schlieren image represents line-of-sight integrated values of the refractive-index gradient of the axisymmetric 3D flow. Hence, the vertical line represents the occurrence of a vertical part of the VRES in the circumferential direction. 

Figure~\ref{fig:Jet_evolution_Def}(e) shows the moment when the Mach disk is located at its maximum axial position of about x/D $\approx$1.17. While the vortex ring propagates further downstream, a triple shock system becomes apparent. The shock system consist of the barrel shock, the Mach disk and the reflected shock. The reflection of the barrel shock from the jet centerline (axis of symmetry) must be, by its nature, a Mach reflection \cite{hornung1986regular}. The corresponding triple point and the slipstream are shown in figure~\ref{fig:Jet_evolution_Def}(e). The slipstream occurs because of the velocity mismatch between the region downstream of the reflected shock and downstream of the Mach disk. As seen in figure~\ref{fig:Jet_evolution_Def}(d-j), a number of counter-rotating vortex rings (CRVRs) occur, resembling the classical slipstream flow features \cite{edgington2014underexpanded}. These vortex rings are generated by the Kelvin-Helmholtz-instability of the shear layer along the slipstream~\citep{ishii1999experimental}.

A number of subsequent compression waves are visible at the time t = 0.498 ms, as shown in figure~\ref{fig:Jet_evolution_Def}(g). They are most likely generated due to reflection of pressure waves on the boundaries in the PDC:  both, the orifices within the DDT section and the cross-section contraction between the DDT section and the exhaust tube.

Figure~\ref{fig:Jet_evolution_Def}(g-j) shows the later development of the CRVRs during the blow down phase and further relevant flow features. The truncated cone shape of these vortex rings represents the evolution of the triple point since this point is the origin of the KH instabilites leading to CRVRs. As the Mach disk becomes smaller with time, the triple point moves further inward. This leads to the truncated cone shape distribution of the CRVRs. The CRVRs remain downstream of the VRES for at least $4 D$. They move continuously away from the jet centerline in the radial direction. Their truncated conical shape (figure~\ref{fig:Jet_evolution_Def}(g)) is due to the inclination of the shear layer and the trajectory of the triple point.  Eventually, the CRVRs move around the vortex ring and reappear \tb{in the field of view,} upstream of the vortex ring (Figure~\ref{fig:Jet_evolution_Def}(j)). The propagation of the CRVRs around the main vortex ring occurs shortly after the vortex ring separates from the trailing jet. 

The vortex ring separation, commonly referred to as pinch-off, takes place between t = 0.498 and 0.598 ms. (figure~\ref{fig:Jet_evolution_Def}(g)-(h)). While  the vortex ring is still attached to its trailing edge at t = 0.498 ms (figure~\ref{fig:Jet_evolution_Def}(g)), the VRES is separated from the reflected shock at 0.598 ms (figure~\ref{fig:Jet_evolution_Def}(h)). As the vortex ring moves further downstream, the first shock cells appear, resembling the flow field of a steady underexpanded jet (figure~\ref{fig:Jet_evolution_Def}(i)-(k))~\citep{Franquet:0gi}. As more and more combustion products leave the tube, the shock cells become less visible and disappear behind the high-density gradients of the burnt gas (figure~\ref{fig:Jet_evolution_Def}(l)). 

\subsection{\textbf{Medium fill-fraction (\Detpart)}}

As described in section~\ref{sec:Operating conditions}, the fill time for the \Detpart case is longer than the \Def case, resulting in most of the tube containing combustible mixture at time of ignition. Therefore, a successful transition to detonation takes place within the DDT section. The resulting detonation wave travels through the combustible mixture until it reaches the intersection of the reactive mixture with air. The detonation wave is then transmitted as a shock. This shock propagates further through the  air-filled section of the tube, as indicated in figure~\ref{fig:cases_and_valve_time_line}(c). A non-reactive shock exits the tube, in the same manner as in the \Def case. However, this shock is much stronger due to the detonation process. Consequently, the incident shock Mach number   M = 2.53 for the \Detpart case is  considerably greater than M = 1.65 of the \Def case.

\begin{figure}
	\centering
	\includegraphics[width=0.9\textwidth]{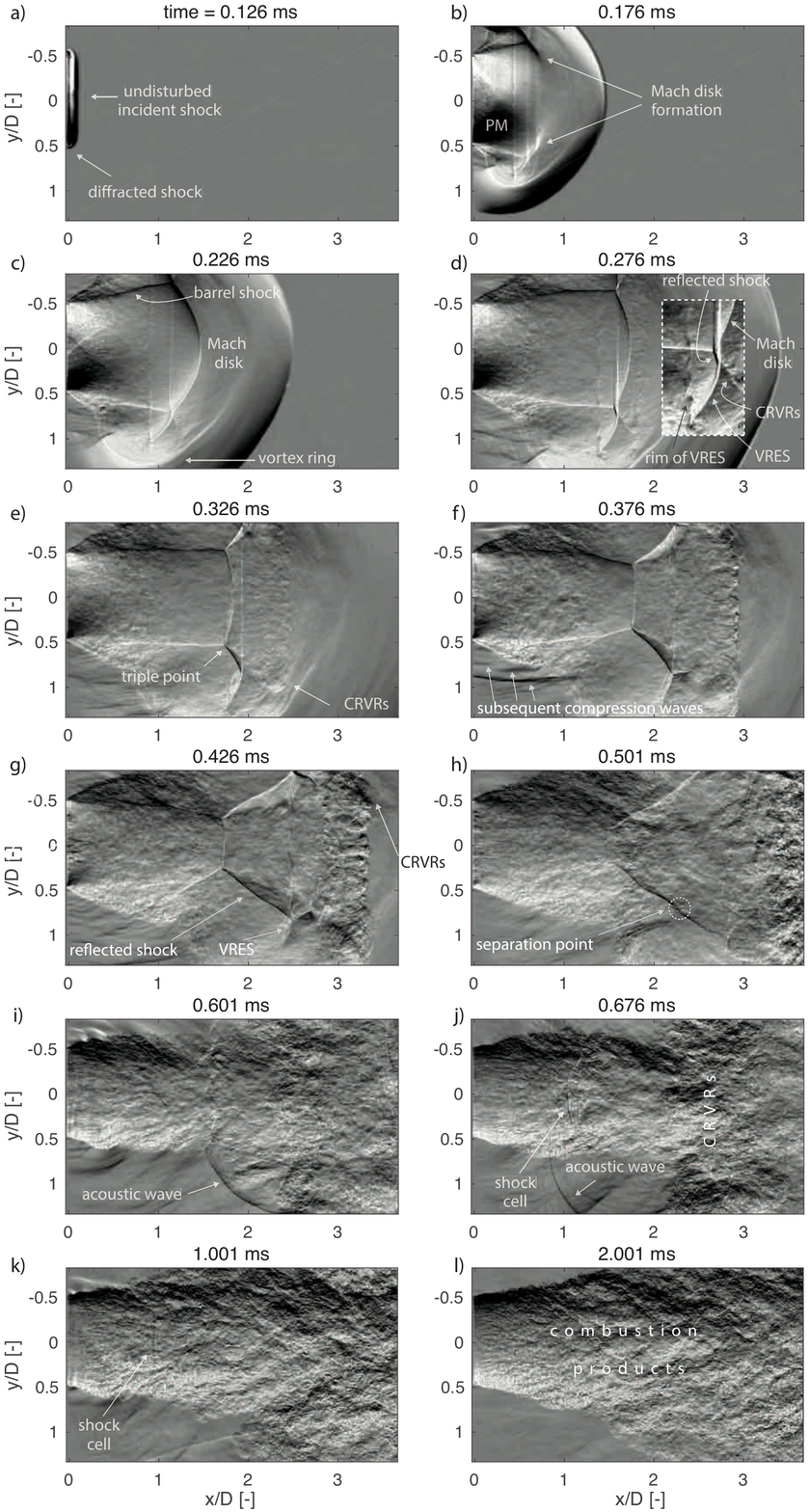}
	\caption{Details of the flow features while the shock undergoes a diffraction. The jet evolution is visible until the combustion products dominate the measurement area. The time interval between the images is 0.025 ms. The given time above each image corresponds to the time elapsed after the shock passes the  pressure probe next to the exhaust tube outlet. Case: \Detpart. }
	\label{fig:Jet_evolution_Det_part}
\end{figure}

A series of $\frac{\partial \rho }{\partial y}$  schlieren images is presented for the \Detpart case in figure~\ref{fig:Jet_evolution_Det_part}. The first image at t = 0.126 \textit{ms} depicts the moment just after the shock leaves the tube. Both the diffracted and undisturbed incident shocks are visible. By the time t = 0.176 \textit{ms } (figure~\ref{fig:Jet_evolution_Def}(b)) the incident shock is fully diffracted, indicated by its curved shape. The Prandtl-Meyer expansion waves located between the barrel shock and the jet centerline are clearly visible. The formation of the Mach disk can be observed from the  figure~\ref{fig:Jet_evolution_Def}(b)-(c). 

The flow features observable in figure~\ref{fig:Jet_evolution_Det_part}(a-d) for the \Detpart case are essentially the same as for the \Def case (figure~\ref{fig:Jet_evolution_Def}(a-d)), which were discussed in the previous section. However, with increasing incident shock Mach number, the shocks produced in the jet are   stronger as is their influence on   the jet evolution ~\cite{ishii1999experimental};  both the size and the axial distance of the Mach disk are larger when compared  to the \Def case (figure~\ref{fig:Jet_evolution_Def}(c) and~\ref{fig:Jet_evolution_Det_part}(c)). In addition, the shape of the barrel shock is less conical but more barrel like, which is characteristic of highly underexpanded jets \cite{Franquet:0gi}. In figure~\ref{fig:Jet_evolution_Det_part}(d) a pronounced wavy line at the rear of the VRES appears. This line represents the impingement point of the shock on the jet boundary~\cite{kleine2010time}. This wavy line is one of several flow features more pronounced as compared to the \Def case.

The later stage of the jet evolution, shown in figure~\ref{fig:Jet_evolution_Det_part}(e-i) is also consistent with the evolution of the flow for the \Def case.  The Mach disk becomes smaller while its axial distance to the tube outlet decreases with time. While the vortex ring propagates further downstream, it separates from the trailing jet. Figure~\ref{fig:Jet_evolution_Det_part}(e) depicts the moment as the elongated reflected shock is about to detach from the VRES.  The rear part of the reflected shock separates at its intersection with the shear layer and propagates as an acoustic wave toward the tube exit (figure~\ref{fig:Jet_evolution_Det_part}(h-j)). By the time t = 0.676~ms the vortex ring is separated from the trailing jet and the CRVRs upstream of the vortex ring appear. 

In the same manner as for the \Def case, the high-density gradient combustion products mask the shock cell structure at later times. Due to the large fill-fraction this occlusion occurs sooner in the schlieren images as compared to the \Def case. Hence, in contrast to the \Def case the diamond shape shock structure is barely visible. However, The vertical part of the first shock cell is notable in figures~\ref{fig:Jet_evolution_Det_part}(j)-(k). 

The appearance of combustion products is confirmed by the signal of the ion probe mounted close to the tube exit. It detects the passage of combustion wave at \mbox{x/D = - 4}  at the latest from t = 0.55~ms on. This correlates very well with the observations from the schlieren images and it confirms the assumption of the high-density gradient combustion products masking the underlying shocks. The image in figure~\ref{fig:Jet_evolution_Det_part}(i) depicts the exhaust flow at t = 2.001~ms. At this time the high-density gradient exhaust flow dominates the entire schlieren image. 

As presented above, the dynamic flow features observed for the \Def and \Detpart operating conditions are qualitatively similar. However, for the \Detpart case the shocks are significantly stronger, as are the flow structures generated by or associated with these shocks.  Moreover, the flow characteristics at both operating conditions correspond very well with those reported in the literature for the exhaust flow of a shock tube and for a compressible starting jet \cite{kleine2010time,ishii1999experimental,dora2014role,fernandez2017compressible,radulescu2007transient,kleine2010timejournal}. \tb{This is not surprisings, as the early phase of the exhaust flow for both the partially filled PDC and a shock tube is dominated by a transient incident shock wave propagating toward the tube exit. However, there is a significant distinction with regard to the evolution of the flow features. In contrast to the open-ended shock tube, the flow features in the PDC exhaust weaken at an early stage. This is in particular noticeable in the evolution of the Mach disk. In the classical open-ended shock tube flow, the Mach disk converge continuously to its steady size and position \cite{hamzehloo2014large,ishii1999experimental}. However, the Mach disk in the PDE exhaust shrinks and moves toward the tube exit soon after its formation is completed, regardless of the leading shock Mach number. This is believed to be due to the presence of rarefaction waves attached to the leading shock for both \Detpart and \Def cases, as discussed in section \ref{sec:Operating conditions}. Consequently, once the shock wave leaves the tube, the pressure declines at the PDC exit. Hence, the less underexpanded flow results in weaker flow features over time. Nevertheless, except for the strength of the flow features, the overall initial PDC exhaust flow dynamics remain qualitatively similar to the one of the open-ended shock tube.}

\FloatBarrier

\subsection{\textbf{Maximum fill-fraction (\Detfull)}} \label{chap:jet_det_full}
\begin{figure}
	\centering	
	\includegraphics[width=0.7\textwidth]{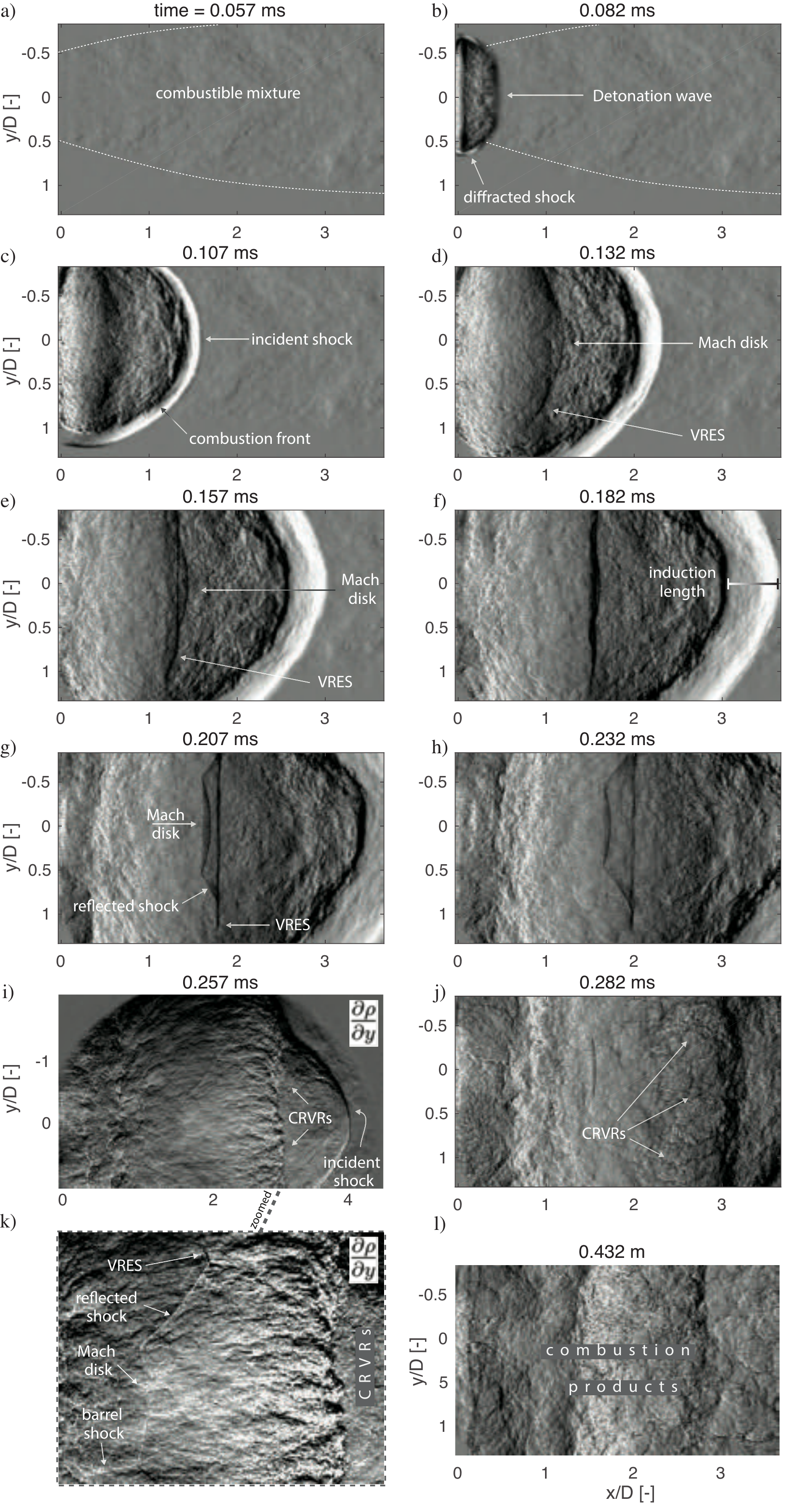}
	\caption{The evolution of the Mach disk and the vortex ring embedded shock is shown as a detonation wave exits an overfilled tube. The embedded shock overtakes the Mach disk about 0.15 ms after the detonation wave emerges from the tube. These shocks becomes less visible as they become weaker and simultaneously more detonation products come out of the tube. Also the presence of CRVRs can be seen in the combustion region.  Case: \Detfull.  }
	\label{fig:Jet_evolution_Det_full}
\end{figure}

When the tube is overfilled with combustible mixture, a detonation wave propagates the entire length of the exhaust tube. A time sequence of schlieren images at the outlet of the PDC is presented in figure~\ref{fig:Jet_evolution_Det_full}. The figures~\ref{fig:Jet_evolution_Det_full}(a) - (j) correspond to a time sequence with time increments of 25 \textmu s. Figure~\ref{fig:Jet_evolution_Det_full}(k) is a close-up view of the figure~\ref{fig:Jet_evolution_Det_full}(i), which shows a larger section of the exhaust tube. All images in this figure represents \drhodx schlieren images with the exception of figures~\ref{fig:Jet_evolution_Det_full}(i) and (k), which represents \drhody  images. 

The outflow of combustible mixture from the tube can be seen in the schlieren image (figure~\ref{fig:Jet_evolution_Det_full}(a)) due to the strong density gradients between the ambient air and the fuel. At t = 0.082  the detonation  has already passed the exit of the exhaust tube  (figure~\ref{fig:Jet_evolution_Det_full}(b)).  The combustion region is characterized by a dark and noisy pattern. This pattern occurs as a result of density gradients in a very small length scale. In the vicinity of the jet core, the undisturbed part of the wave propagates as a detonation wave, while the outer part of the incident shock diffracts at the corner (figure~\ref{fig:Jet_evolution_Det_full}(b)). The diffracted part of the wave is characterized  by a small, bright region. This region corresponds to the zone between the incident shock and the combustion front. The jet boundary of the reactive mixture marks the intersection point between the detonation wave and diffracted shock with no combustion front right behind it. By the time t = 0.107 ms  the incident shock is already fully separated from the combustion region along the entire wave front (figure~\ref{fig:Jet_evolution_Det_full}(c). Hence, the detonation wave fails shortly after it exits the tube. 

The Mach disk and the VRES are initially entirely obscured by the combustion products, but become apparent  (figures~\ref{fig:Jet_evolution_Det_full}(d)-(e)), as the jet expands further and the strength of the shocks increases. Although the vortex ring is not observable due to the combustion products, the presence of the VRES indicates the presence of a vortex ring. Similar to the exhaust flow of the \Detpart and \Def cases the VRES overtakes the Mach disk (figures~\ref{fig:Jet_evolution_Det_full}(g)-(h)).

The evolution of the flow features shown in figure~\ref{fig:Jet_evolution_Det_full} and its similarity to those of the other operating conditions lead to the conclusion that the exhaust flow of the PDC for \mbox{ff \textsubscript{Def} \textgreater\ 1}  undergoes qualitatively the same fluid dynamic development as it does for \mbox{ff \textsubscript{Def} \textless\ 1}. Therefore, we assume the presence of a Mach reflection for an overfilled PDC in the same manner as for the cases with a smaller fill-fraction. The Mach reflection consists of the reflected shock, the Mach disk and the barrel shock. However, the latter is not visible in the \drhodx images (figure~\ref{fig:Jet_evolution_Det_full}(g)-(h)). This is again due to the combustion products masking the underlying barrel shock, as its pressure gradient in the axial direction are relatively small. The barrel shock becomes visible in the \drhody image as shown in figure~\ref{fig:Jet_evolution_Det_full}(i) and is marked in the corresponding close-up view in figure~\ref{fig:Jet_evolution_Det_full}(k). 

As shown in  figure~\ref{fig:Jet_evolution_Det_full}(i), CRVRs appear more distinctly in the \drhody schlieren image in the region x/D $\approx$ 2.5 to 3.5.  The CRVRs are also notable in the  \drhodx image as a dark region elongated in vertical direction in figure~\ref{fig:Jet_evolution_Det_full}(j). With time the combustion products dominate the image more and more( figure~\ref{fig:Jet_evolution_Det_full}(l)). Therefore, the further analysis of the flow development based on schlieren images becomes unfeasible. 

\begin{figure}
	\centering
	\includegraphics[width=0.9\textwidth]{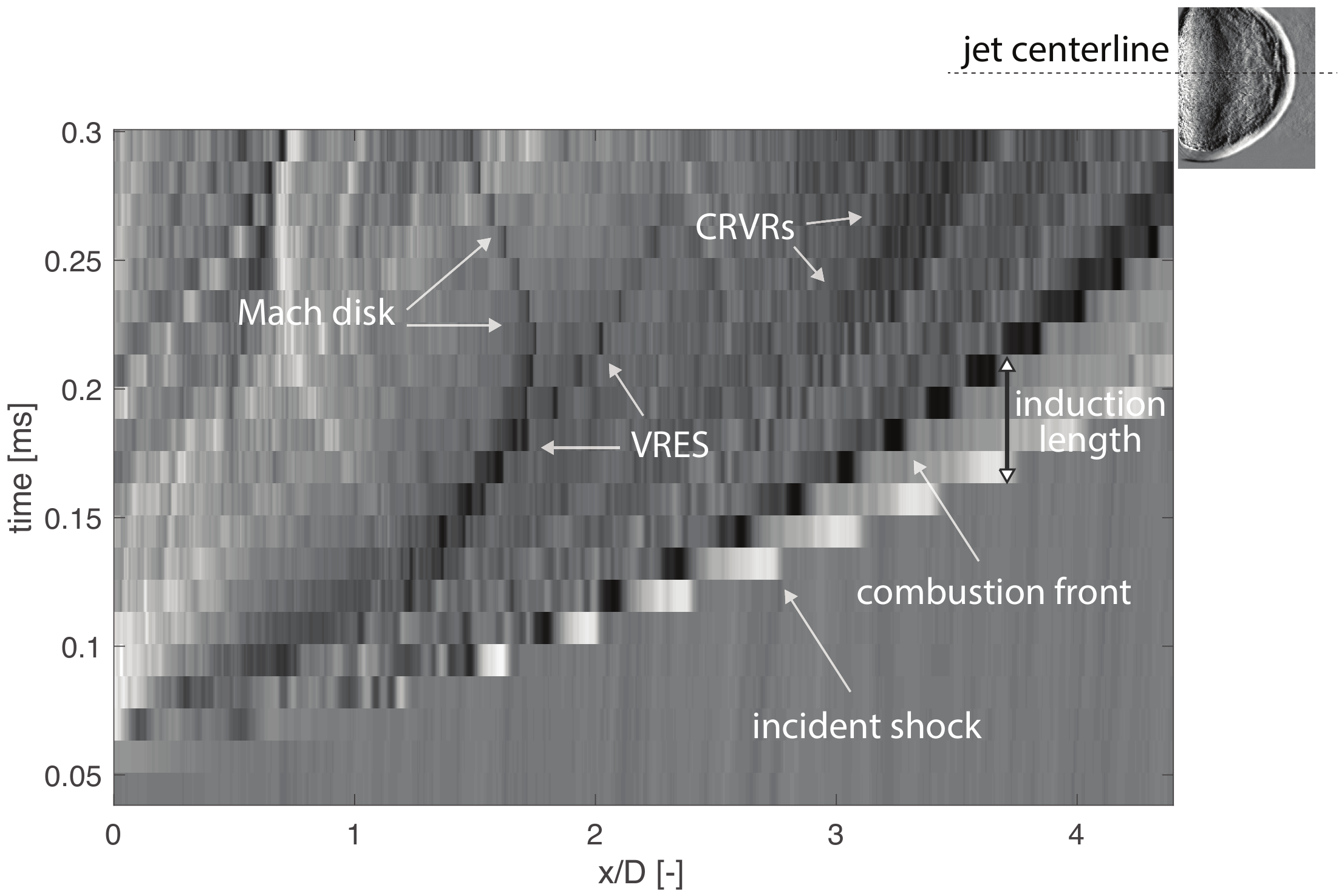}
	\caption{x-t diagram showing the evolution of the main flow features of the diffracting detonation along the jet centerline. The x-axis is normalized by the tube diameter, while the y-axis shows the time from 0 to 0.3 ms. Case: \Detfull.  }
	\label{fig:x_t_Det_full}
\end{figure}

The spatio-temporal evolution of the essential flow features can be evaluated based on an x-t diagram. Figure~\ref{fig:x_t_Det_full} shows an x-t diagram derived from the schlieren images recorded at 80 kHz along the jet centerline. To reduce the noise, the pixel intensity is averaged in vertical direction for 30 pixels, which corresponds to $0.15 D$.  The combustion front decays at a faster rate than the incident shock, which results in a continuous growth of the induction length.  At  t $\approx $  0.2 ms the Mach disk reaches its maximum axial distance and recedes backward toward the tube exit. The vertical line of the VRES  indicates the propagation path of the vortex ring. Also the CRVRs upstream of the vortex ring are notable in the x-t diagram, indicating that these vortex rings move around the primary vortex ring. With the exception of the combustion front, the dynamic evolution of all flow features corresponds qualitatively very well with the under-filled operating conditions, as discussed in the previous chapters. 

To summarize, the above discussion emphasizes the similarity of the main flow features for all three operating conditions. This is not only true for the presence of these features but also for their dynamical evolution.

\section{Influence of Fill Fraction on the Full PDC cycle }\label{sec:Jet_evolution_full_cycle}
After the discussion of the initial development of the exhaust flow for different operating conditions, we now address the full cycle of the PDC. For this purpose the PIV results are utilized  and the impact of fill-fraction on the full cycle of the exhaust flow is examined. The cases  \Def and  \Detfull, representing the minimum and maximum  fill-fractions, are investigated.

\subsection{Flow snapshots during blow down} 
\begin{figure}
 	\centering
 	\includegraphics[width=0.8\textwidth]{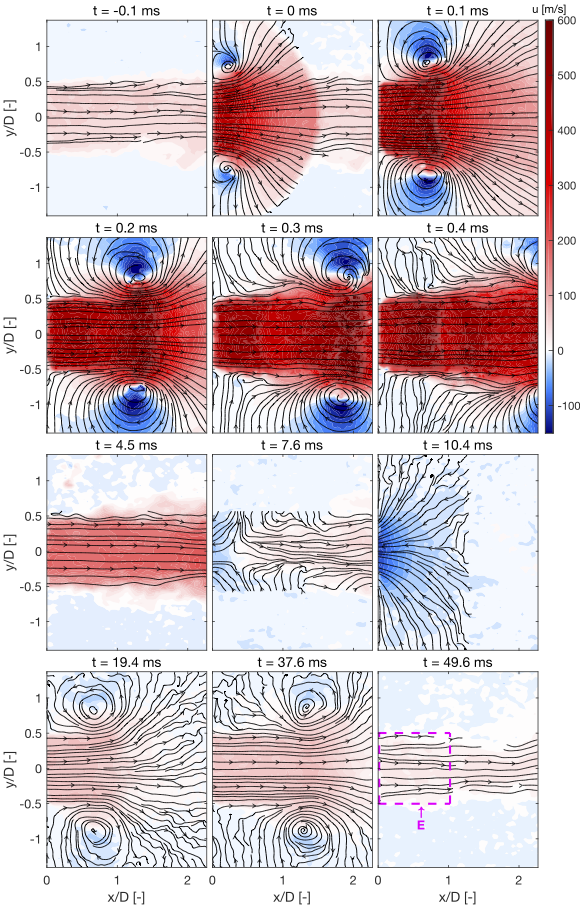}
 	\caption{Contours of axial velocity of the exhaust flow  at 12 time steps showing the evolution of the flow for the \Def case. \tb{ Streamlines calculated from instantaneous velocity fields are superimposed on contours to indicate flow direction.}}
 	\label{fig:piv_vectorplots}
 \end{figure}

\begin{figure}
 	\centering
 	\includegraphics[width=0.8\textwidth]{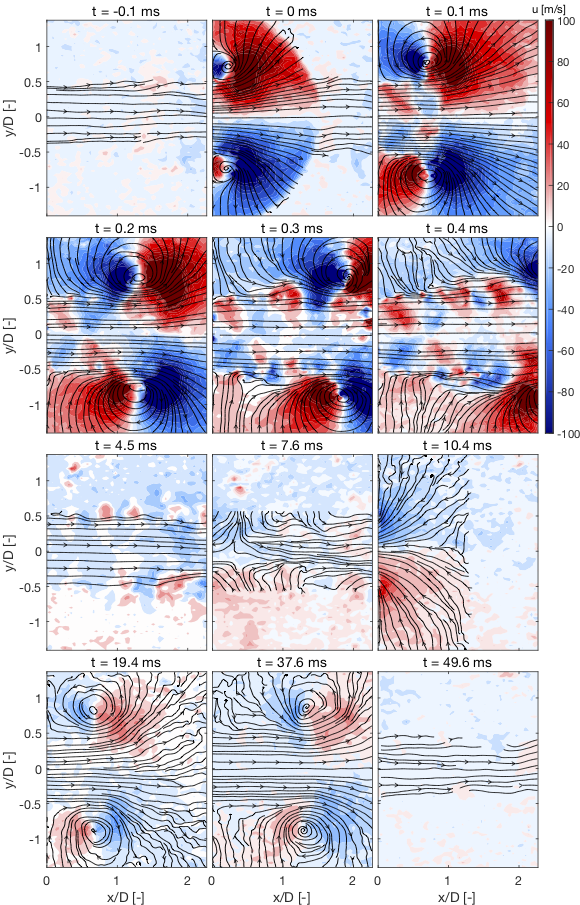}
 	\caption{Contours of radial velocity of the exhaust flow  at 12 time steps showing the evolution of the flow for the \Def case. \tb{ Streamlines calculated from instantaneous velocity fields are superimposed on contours to indicate flow direction.}}
 	\label{fig:piv_vectorplots_v}
 \end{figure}

Figures~\ref{fig:piv_vectorplots} and~\ref{fig:piv_vectorplots_v} show a series of contour plots of the instantaneous axial and radial velocity for the Def case. The first six consecutive plots~(a)-(f) in figures~\ref{fig:piv_vectorplots} and~\ref{fig:piv_vectorplots_v} exhibit the initial stage of the flow evolution. As mentioned in section~\ref{sec:exp_setup} the PDC is purged with air before the combustion takes place. Figure~\ref{fig:piv_vectorplots}(a) shows the corresponding purge flow jet, just before the incident shock arrives at the tube exit. Figure~\ref{fig:piv_vectorplots}(b) depicts the moment, when the incident shock is fully diffracted outside of the tube. The corresponding schlieren image is shown in figure~\ref{fig:Jet_evolution_Def}(b). The maximum velocity occurs where the flow expands through the PM-expansion fan. The negative axial velocity (dark blue region) indicates the presence of a vortex ring. For the subsequent discussion we set the clock to time t = 0 at the instance shown in Figure~\ref{fig:piv_vectorplots}(b), when the incident shock appears for the first time in the measurement domain. Figure~\ref{fig:piv_vectorplots}(c) shows the exhaust flow 0.1 ms later, which corresponds to the schlieren image shown in figure~\ref{fig:Jet_evolution_Def}(d). Compared to figure~\ref{fig:piv_vectorplots}(b) the jet is now further expanded and the vortex ring is larger. As the vortex ring propagates further downstream (figure~\ref{fig:piv_vectorplots}(d)-(f)), the jet close to the tube exit begins to exhibit the shock structures characteristic of an underexpanded jet. The radial velocity presented in figure~\ref{fig:piv_vectorplots_v}(d)-(f) shows the typical velocity field for a shock cell structure within the supersonic jet ($-0.5<y/D<0.5$).  The shock cell structure is characterized by regions of positive and negative radial velocity, as the flow is redirected in transverse direction by the oblique shocks. 

When the leading shock wave leaves the tube, a set of rarefaction waves are generated, which propagate back into the tube. Therefore, the pressure inside the tube decreases, and consequently the jet velocity at the tube exit decreases. Eventually, the pressure drops sufficiently low that the supersonic underexpanded jet  becomes a subsonic jet. From figure~\ref{fig:piv_vectorplots_v}(g) it seems that at t = 4.5 ms the supersonic underexpanded jet has already transformed to a subsonic jet, as no shock cell structure can be observed. While the pressure inside the tube decays further, the jet velocity decreases accordingly until the flow direction reverses.  Figure~\ref{fig:piv_vectorplots}(h) shows the moment when the fluid close to the tube exit flows back toward the tube. Figure~\ref{fig:piv_vectorplots}(i), shows a high momentum reverse flow with a minimum axial velocity of -109.2 m/s, which takes place at the outlet. The first suction phase lasts about ten milliseconds. Figure \mbox{\ref{fig:piv_vectorplots}(j)-(k)} shows two vortex rings at t = 37.6 and t = 49.6 ms, which indicate two additional exhaust phases. Figure~\ref{fig:piv_vectorplots}(l) shows the purge jet, as the blow down process is finished. 

\subsection{Global flow quantities during blow down}
To get an impression of the overall flow fluctuations that occur during the PDC cycle, we extract centerline global flow quantities from the PIV data.  The first quantity, termed  $u_{\mathrm{min}/\mathrm{max}}$ represents the {\itshape local} velocity minima or maxima, while the second quantity $\dot{V}$ represents the {\itshape global} (radially integrated) streamwise volume flux close to the tube exit. With these two quantities  the local extreme events and the overall global dynamics of the flow field at the exhaust can be quantified.

The  volume flux at the tube exit is defined as  $\dot{V}(t)=\oint_Au_x(t)\mathrm{d}A$, where $u_x$ is the instantaneous streamwise velocity component and $A$ is the tube cross-sectional area. Since the PIV data is only available in a plane perpendicular to the cross-sectional area, the integral is conducted only in radial direction assuming an axisymmetric flow field. As seen from the  snapshots  shown in fig.~\ref{fig:piv_vectorplots}, this assumption is valid. Moreover, PIV is not available exactly at the tube outlet and the volume flux is determined at the most reliable upstream measurement location of $x/D=0.3$.  

Regarding the local measure $u_{\mathrm{min}/\mathrm{max}}$ of the exhaust flow velocity, the definition is less straight-forward. The purpose of this quantity is to reveal the highest absolute velocity occurring in the flow field during the blow-down and  suction phases.
For this purpose we define a region E near the tube exit, where this quantity is evaluated (see dashed rectangular in figure~\ref{fig:piv_vectorplots}(l)). We then evaluate the average flow direction from spatially averaging the vectors within this area. Thereafter we search for the highest velocity in this direction in area E and assign this value to the quantity $u_{\mathrm{min}/\mathrm{max}}$.  In practise, instead of taking the highest velocity, we take an average of the five highest velocities to  minimize the impact of outliers. In that sense, this quantity represents the local maximum axial velocity in the vicinity of the tube exit, while accounting for the flow direction. It complements the global measure of the axial volume flux.

\begin{figure}
	\centering
	\includegraphics[width=1\textwidth]{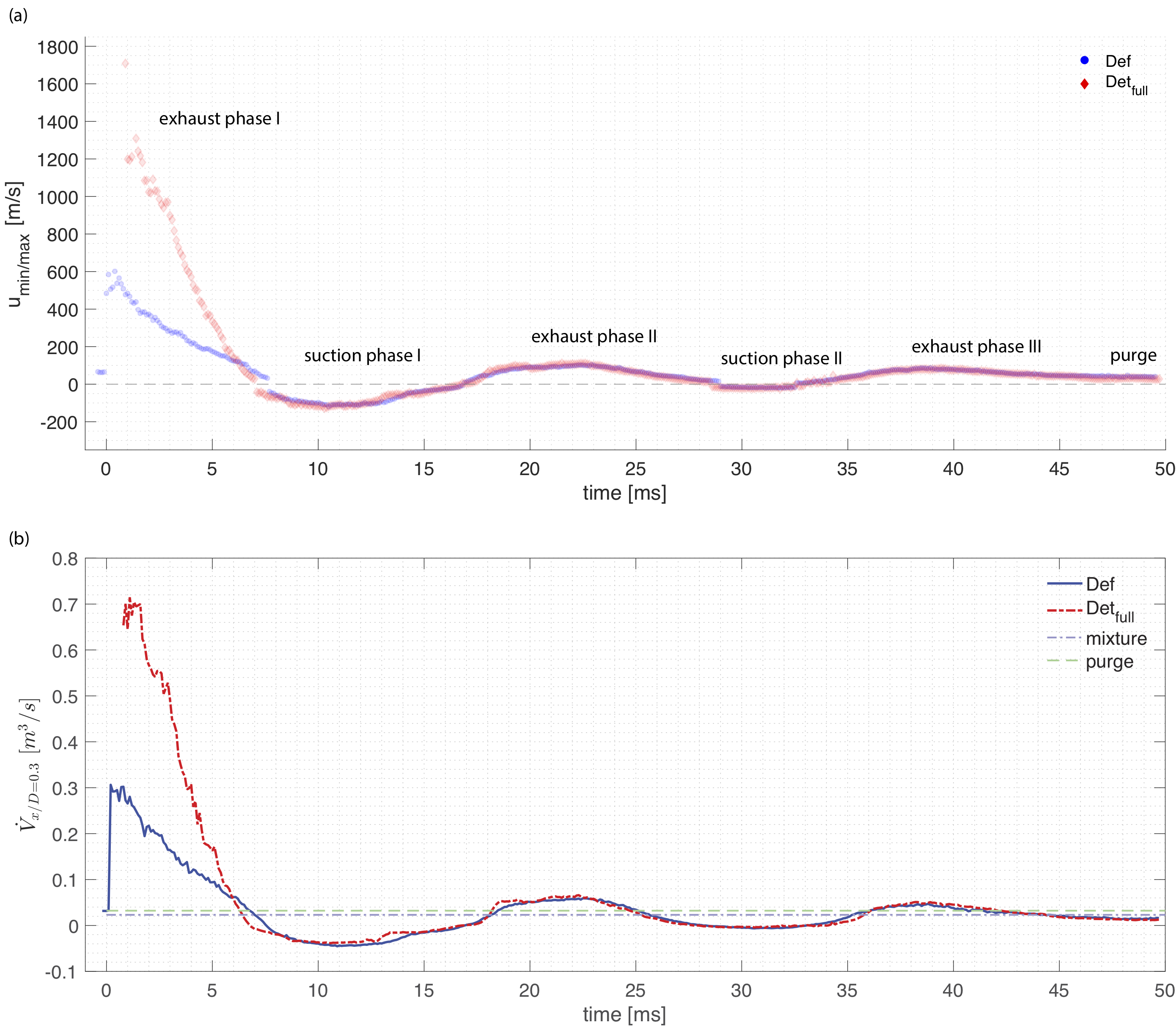}
	\caption{(a) The axial velocity  $u_{\mathrm{min}/\mathrm{max}}$ and (b) the volume flux at x/D = 0.3 over time representing the exhaust and suction phases of the PDC exhaust flow for \Detfull and Def cases. }
	\label{fig:piv_scatter}
\end{figure}

Figure~\ref{fig:piv_scatter}(a) presents the quantity  $u_{\mathrm{min}/\mathrm{max}}$ for both Def and \Detfull cases for the first 50 ms of the PDC cycle. We start evaluating the Def case before proceeding further with the \Detfull case. After the shock wave leaves the tube, the velocity increases up to 483 m/s at t = 0. The further expansion of the jet caused by the expansion waves (figure~\ref{fig:piv_vectorplots}(f)) leads to a maximum velocity of 601 m/s at t = 0.4 ms. The oscillation of $u_{\mathrm{min}/\mathrm{max}}$  up to t = 0.5 ms is mainly due to the presence of shock cells (figure~\ref{fig:piv_vectorplots_v}(c)-(f)). From t = 0.5 ms to 7.6 ms the velocity decreases gradually (figure~\ref{fig:piv_scatter}(a)). At t = 7.7 ms the axial velocity is negative,  indicating the presence of a reverse flow. The maximum back flow velocity of  -118 m/s occurs at \mbox{t = 10.5 ms}. The first suction phase lasts up to t = 16.9 ms. A secondary exhaust phase takes place from t = 17 to 28.9 ms (exhaust phase II) with a maximum velocity of 104.6 m/s. This is followed by a secondary suction phase from t = 29 to 32.5 ms. A third exhaust phase is notable from t = 32.6 which is again followed by an decreasing velocity phase. The several exhaust and suction phases are caused due to the compression and expansion waves propagating inside the PDC. These waves reflect at the interface between the combustible mixture and the air within the DDT section (figure~\ref{fig:cases_and_valve_time_line}(c)), the cross-section contraction, the close end and the open end of the tube. 

\begin{figure}
 	\centering
 	\includegraphics[width=1\textwidth]{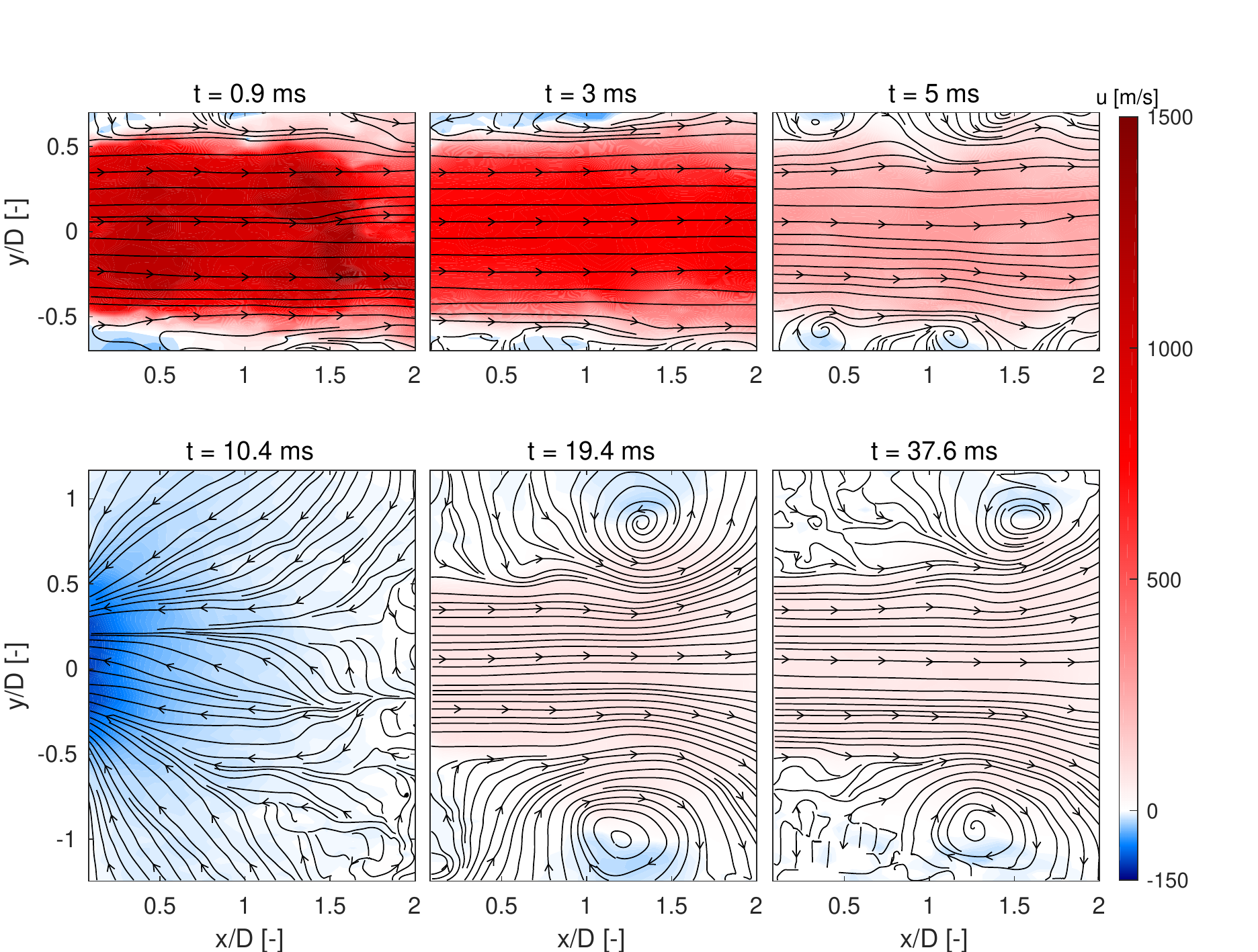}
 	\caption{Axial velocity fields of the exhaust flow at 6 time steps  showing the evolution of the flow for the \Detfull case.\tb{   Streamlines calculated from instantaneous velocity fields are superimposed on contours to indicate flow direction.}}
 	\label{fig:piv_vectorplots_Det}
 \end{figure}
 
The impact of the fill-fraction on the evolution of the exhaust flow is estimated by comparing the axial velocity and the volume flux at the tube exit of the \Detfull case to the Def case. As seen in figure~\ref{fig:piv_scatter}(a), the \Detfull case also features three exhaust phases and two suction phases. In comparison to the Def case, the axial velocity in the first exhaust phase is significantly higher, which is caused by the detonation wave propagating through the entire exhaust tube. Due to difficulties with the seeding of the flow just upstream the detonation wave, the first accurate measurement point occurs at t = 0.9 ms. The velocity at this time is the maximum measured axial velocity of the exhaust flow \mbox{($u_{\mathrm{min}/\mathrm{max}}$ = 1708 m/s)}. From figure~\ref{fig:piv_scatter}, it is clear that in the first exhaust phase the flow decelerates at a higher rate when compared to the Def case. Figure~\ref{fig:piv_vectorplots_Det}(a)-(c) shows the corresponding axial velocity contour plots of the jet.  The first suction phase begins at \mbox{t = 7.1 ms} as shown in figure~\ref{fig:piv_scatter}. From here on the jet evolution is very similar to the Def case. The high momentum reverse flow and the subsequent vortex rings of the second and third exhaust phases are presented in figure~\ref{fig:piv_vectorplots_Det}(d)-(f), respectively. These plots show the velocity field of the \Detfull case for the same moments (t = 10.4, 19.4 and 37.6 ms) as figure~\ref{fig:piv_vectorplots}(i)-(k) for the \Def case. The almost identical flow fields emphasize the similarity of the exhaust flow, once the first suction phase begins. 

Figure~\ref{fig:piv_scatter}(b) shows the global volume flux $\dot{V}$ evaluated at $x/D = 0.3$ for both \Detfull and \Def cases. The horizontal lines in the figure further indicate the volume flux determined from the pre-set values of the air and hydrogen flow rates for the mixture and purge phases. As seen for the Def case, the pre-set value for the purge phase agree very well the time $t=0$, which justifies the accuracy of the PIV results and the determination of $\dot{V}$. Once the  shock wave exits the tube the volume flux increases abruptly from 0.032 $m^3/s$ at t = 0 to its maximum value of 0.31 $m^3/s$ at t = 0.2 ms for the \Def case and more than twice this value for the \Detfull case. The relative difference between the two is very similar to the one observed for the local streamwise velocities discussed before. Moreover, the global volume flux shows a very similar oscillatory behaviour than the local velocity measure. It is further evident that the oscillations asymptotically converge to the pre-set volume flow (horizontal line).   

These results suggests that for the current PDC design the fill-fraction impacts only the first exhaust phase in regard to the exhaust velocity. The subsequent suction and exhaust phases are very similar in terms of timing, velocity magnitude and the overall flow topology. \tb{ The exhaust flow after the first exhaust phase is assumed to be mainly controlled by the rarefaction and compression waves travelling along the tube. Interestingly, the pressure measured close to the tube outlet shows similar arriving time and amplitude for the subsequent expansion and compression waves for a number of different fill fractions. The mechanism responsible for this nearly identical wave dynamics, however, remains an open question.}

\section{Conclusion}

In this study, the exhaust flow of a PDC was investigated using high-speed schlieren and PIV. To conduct reliable PIV measurements for theses strongly  shock-driven flows, the relaxation time of different PIV seeding particles were evaluated in a preliminary study based on PIV of a highly underexpanded steady jet. These studies suggest that TiO\textsubscript{2} particles are most suitable in terms of time lag and particle dispersion. These particles were then used to conduct PIV in the PDC exhaust flow, for three  different PDC fill-fractions.  To the authors knowledge, this is the first time that high-speed PIV is used to resolve the full cycle of a PDC. The acquired flow data combines very well with the  high-resolution schlieren images presented along this study.    

The high-speed schlieren images reveal the initial evolution of the flow features in detail. For the partially filled tube, the exhaust flow corresponds to the flow field of a classical open-ended shock tube. The incident shock diffracts as it exits the tube. The flow is characterized by a vortex ring and a secondary shock system. A number of counter-rotating-vortex-rings emerge from the shear layer of a slipstream, which originates from the triple point of a Mach reflection.  These vortex rings propagate around the main vortex ring shortly after the main vortex ring separates from the trailing jet. After the pinch-off process, an underexpanded jet evolves, which is characterized by its typical shock diamond structure. As more and more combustion products leave the tube, the schlieren images become dominated by high-density gradient combustion products. Very similar flow features are also found for the cases with higher fill-fraction where a detonation wave occurs. 
Based on the schlieren images, the same flow dynamics are observed but the strength and size of the flow features has increased, as the velocity of the leading shock increases. In this situation, however, many details about the shock structures are blurred in the schlieren images due to the combustion products arriving shortly after the leading shock wave. 

 The full cycle of the exhaust flow is investigated based on the PIV data acquired at the tube exit. The high-speed PIV results show that the initial exhaust phase is followed by an even longer suction phase. A subsequent second exhaust-suction phase occurs, which is followed by a third exhaust phase. These exhaust and suction phases are caused by a number of compression and expansion waves propagating inside the PDC. These results are very well suited for validation of low-dimensional numerical schemes developed for PDC development \cite{Nadolski2019}. 
Comparing the PIV results for the partially and overfilled operating condition shows that the fill-fraction only affects the first exhaust phase in respect to the local axial velocity and global volume flux. The subsequent suction and exhaust phases are surprisingly similar in terms of timing and velocity. The fact that the local flow features determined from schlieren also remain similar, lead to the conclusion that the exhaust flow of the PDC for the overfilled configuration undergoes the same fluid dynamic development as it does for a partially filled case.  Although, with increasing fill-fraction the combustion products occur earlier, the nature of the exhaust flow remains unchanged.

The characterization of the dynamic evolution of the PDC exhaust flow can be used for various purposes. The detailed description in this work concerning the evolution of the various flow features and their interaction with each other  helps to gain a better understanding of the exhaust flow of a detonation or shock tube. Moreover, from the application point of view for the PDC, these results support the design and optimization process for the coupling of the PDC with a turbine. Furthermore, the results can be used as a benchmark for validation of computational simulations.

\FloatBarrier

\section*{Acknowledgments}
The authors gratefully acknowledge support by the Deutsche Forschungsgemeinschaft (DFG) as part of collaborative research center SFB 1029 "Substantial efficiency increase in gas turbines through direct use of coupled unsteady combustion and flow dynamics" on project C01. The authors also acknowledge the FDX Fluid Dynamix GmbH for providing fluidic oscillators. 

\bibliography{0_AIAA_PDE_main}

\end{document}